\documentclass[aps,prl,a4paper,superscriptaddress,twocolumn,showpacs,amsmath,amssymb]{revtex4}

\usepackage{graphicx,epsfig}
\usepackage[usenames]{color}

\usepackage[english]{babel}

\usepackage{graphicx,epsfig}

\allowdisplaybreaks
\begin{document}

\newcommand{\cau}{\underline{c}^{\phantom{\dagger}}}
\newcommand{\ccu}{\underline{c}^\dagger}

\newcommand{\hilb}{\mathcal{H}}
\newcommand{\hfa}{\hat{f}^{\phantom{\dagger}}}
\newcommand{\hcc}{\hat{c}^\dagger}
\newcommand{\hca}{\hat{c}^{\phantom{\dagger}}}
\newcommand{\hfc}{\hat{f}^\dagger}

\newcommand{\Dp}{\hat{\Delta}_{\text{p}}}
\newcommand{\Dh}{\hat{\Delta}_{\text{h}}}
\newcommand{\Deltap}{[\hat{\Delta}_{\text{p}}]}
\newcommand{\Deltah}{[\hat{\Delta}_{\text{h}}]}
\newcommand{\R}{\mathcal{R}}
\newcommand{\Rh}{\hat{\mathcal{R}}}

\newcommand{\uA}{|\underline{A},Ri\rangle}
\newcommand{\uB}{|\underline{B},R'j\rangle}

\newcommand{\E}{\mathcal{E}}
\newcommand{\G}{\mathcal{G}}
\newcommand{\Lag}{\mathcal{L}}
\newcommand{\M}{\mathcal{M}}
\newcommand{\N}{\mathcal{N}}
\newcommand{\U}{\mathcal{U}}
\newcommand{\F}{\mathcal{F}}
\newcommand{\V}{\mathcal{V}}
\newcommand{\C}{\mathcal{C}}
\newcommand{\I}{\mathcal{I}}
\newcommand{\s}{\sigma}
\newcommand{\up}{\uparrow}
\newcommand{\dw}{\downarrow}
\newcommand{\h}{\hat{H}}
\newcommand{\himp}{\hat{H}_{\text{imp}}}
\newcommand{\g}{\mathcal{G}^{-1}_0}
\newcommand{\D}{\mathcal{D}}
\newcommand{\A}{\mathcal{A}}
\newcommand{\projR}{\hat{\mathcal{P}}_R}
\newcommand{\proj}{\hat{\mathcal{P}}_G}
\newcommand{\K}{\textbf{k}}
\newcommand{\Q}{\textbf{q}}
\newcommand{\T}{\tau_{\ast}}
\newcommand{\io}{i\omega_n}
\newcommand{\eps}{\varepsilon}
\newcommand{\+}{\dag}
\newcommand{\su}{\uparrow}
\newcommand{\giu}{\downarrow}
\newcommand{\0}[1]{\textbf{#1}}
\newcommand{\ca}{c^{\phantom{\dagger}}}
\newcommand{\cc}{c^\dagger}
\newcommand{\Psia}{\Psi^{\phantom{\dagger}}}
\newcommand{\Psic}{\Psi^\dagger}
\newcommand{\aaa}{a^{\phantom{\dagger}}}
\newcommand{\aac}{a^\dagger}
\newcommand{\bba}{b^{\phantom{\dagger}}}
\newcommand{\bbc}{b^\dagger}
\newcommand{\da}{d^{\phantom{\dagger}}}
\newcommand{\dc}{d^\dagger}
\newcommand{\fa}{f^{\phantom{\dagger}}}
\newcommand{\fc}{f^\dagger}
\newcommand{\ha}{h^{\phantom{\dagger}}}
\newcommand{\hc}{h^\dagger}
\newcommand{\be}{\begin{equation}}
\newcommand{\ee}{\end{equation}}
\newcommand{\bea}{\begin{eqnarray}}
\newcommand{\eea}{\end{eqnarray}}
\newcommand{\ba}{\begin{eqnarray*}}
\newcommand{\ea}{\end{eqnarray*}}
\newcommand{\dagga}{{\phantom{\dagger}}}
\newcommand{\bR}{\mathbf{R}}
\newcommand{\bQ}{\mathbf{Q}}
\newcommand{\bq}{\mathbf{q}}
\newcommand{\bqp}{\mathbf{q'}}
\newcommand{\bk}{\mathbf{k}}
\newcommand{\bh}{\mathbf{h}}
\newcommand{\bkp}{\mathbf{k'}}
\newcommand{\bp}{\mathbf{p}}
\newcommand{\bL}{\mathbf{L}}
\newcommand{\bRp}{\mathbf{R'}}
\newcommand{\bx}{\mathbf{x}}
\newcommand{\by}{\mathbf{y}}
\newcommand{\bz}{\mathbf{z}}
\newcommand{\br}{\mathbf{r}}
\newcommand{\Ima}{{\Im m}}
\newcommand{\Rea}{{\Re e}}
\newcommand{\Pj}[2]{|#1\rangle\langle #2|}
\newcommand{\ket}[1]{\vert#1\rangle}
\newcommand{\bra}[1]{\langle#1\vert}
\newcommand{\setof}[1]{\left\{#1\right\}}
\newcommand{\fract}[2]{\frac{\displaystyle #1}{\displaystyle #2}}
\newcommand{\Av}[2]{\langle #1|\,#2\,|#1\rangle}
\newcommand{\av}[1]{\langle #1 \rangle}
\newcommand{\Mel}[3]{\langle #1|#2\,|#3\rangle}
\newcommand{\Avs}[1]{\langle \,#1\,\rangle_0}
\newcommand{\eqn}[1]{(\ref{#1})}
\newcommand{\Tr}{\mathrm{Tr}}

\newcommand{\Vb}{\bar{\mathcal{V}}}
\newcommand{\Vd}{\Delta\mathcal{V}}
\def\P{\hat{\mathcal{P}}_{G}}
\newcommand{\Hqp}{\hat{H}_{\text{qp}}}
\newcommand{\Pd}{\Delta P_{02}}
\def\t{\theta_{02}}
\newcommand{\tb}{\bar{\theta}_{02}}
\newcommand{\td}{\Delta \theta_{02}}
\newcommand{\Rb}{\bar{R}}
\newcommand{\Rd}{\Delta R}
\newcommand{\tlambda}{\tilde{\lambda}}
\newcommand{\tR}{\tilde{\mathcal{R}}}

\def\changemargin#1#2{\list{}{\rightmargin#2\leftmargin#1}\item[]}
\let\endchangemargin=\endlist

\title{Emergent Bloch Excitations in Mott Matter}

\author{Nicola Lanat\`a}
\affiliation{Department of Physics and National High Magnetic Field Laboratory, Florida State University, Tallahassee, Florida 32306, USA}
\author{Tsung-Han Lee}
\affiliation{Department of Physics and National High Magnetic Field Laboratory, Florida State University, Tallahassee, Florida 32306, USA}
\author{Yong-Xin Yao}
\affiliation{Ames Laboratory-U.S. DOE and Department of Physics and Astronomy, Iowa State University, Ames, Iowa IA 50011, USA}
\author{Vladimir Dobrosavljevi\'c}
\affiliation{Department of Physics and National High Magnetic Field Laboratory, Florida State University, Tallahassee, Florida 32306, USA}

\date{\today} 
\pacs{71.27.+a, 71.30.+h,71.10.Hf}


%

\begin{abstract}
We develop a unified theoretical picture for excitations in Mott systems,  portraying both the heavy quasiparticle excitations and the Hubbard bands as features of an emergent Fermi liquid state formed in an extended Hilbert space, which is non-perturbatively connected to the physical system. This observation sheds light on the fact that even the incoherent excitations in strongly correlated matter often display a well defined Bloch character, with pronounced momentum dispersion. 
Furthermore, it indicates that the Mott point can be viewed as a topological transition, where the number of distinct dispersing bands displays a sudden change at the critical point. Our results, obtained from an appropriate variational principle, display also remarkable quantitative accuracy. 
This opens an exciting avenue for fast realistic modeling of strongly correlated materials. 
\end{abstract}

\maketitle

\emph{Introduction:}---
The physical nature of the excited states in strongly interacting quantum systems has long been a subject of much
controversy and debate. Deeper understanding was achieved by Landau, more than half a century
ago~\cite{Landau1}, who realized that in systems of fermions the Pauli principle provides a spectacular simplification. He showed that many properties of Fermi systems can be understood in terms of weakly interacting {\em quasiparticles} (QP), allowing a precise and detailed description of strongly correlated matter.
Modern experiments provide for even more direct evidence of such QP excitations, for example from using angle-resolved photoemission spectroscopy (ARPES)~\cite{arpes} 
or scanning-tunneling microscopy (STM) methods~\cite{STM-PhysRevLett}.

The Fermi liquid paradigm, however, describes only the low energy excitations. At higher energies, the physical properties are often dominated by incoherent processes, which do not conform to the Landau picture. The task to provide a simple and robust theoretical description of such incoherent excitations has therefore emerged as a central challenge of contemporary physics. An intriguing apparent paradox is most evident around the Mott point. Here, ARPES and STM experiments provide often clear evidence of additional well-defined high energy excitations (Hubbard bands) which, while being fairly incoherent, still display relatively well defined Bloch character with pronounced momentum
dispersion, see, e.g., Ref.~\cite{dispersive-Hubbard-FeSe}. 
As a matter of fact, it is often difficult to experimentally even distinguish the Hubbard bands found in Mott insulators from ordinary Bloch bands found at high energy in conventional band insulators. While such behavior can be already numerically reproduced by some modern many-body approximations~\cite{DMFT,dmft_book}, a simple conceptual picture for the apparent Bloch character of such high energy charge excitations is not still available. In particular, variational methods such as the Gutzwiller Approximation (GA)~\cite{Gutzwiller3} 
--- which are often able to reproduce the numerical results in a much simpler semi-analytical fashion --- generally capture only the low-lying QP features on the metallic side, but cannot provide a description of charge excitations around the Mott point and in the insulating regime.

The goal of this Letter is to write an appropriate variational wave function able to capture both the (low energy) QP bands and the (high energy) Hubbard bands, within the same theoretical framework.
A particularly interesting fact 
emerging from our theory is that
many important features of both types of excitations
are encoded in the bare density of states (DOS) of the uncorrelated system and a few renormalization parameters ---
in a similar fashion as for the QP excitations
in Landau theory of Fermi liquids.
This is accomplished, similarly as in many other theories for many-body
systems, see, e.g., Refs.~\cite{AKLT,Vaestrate_review,Ulrich_review}, by enlarging the Hilbert space by
introducing auxiliary "ghost" degrees of freedom.
In particular, this construction sheds light on
the physical origin of the "hidden" Bloch character
of the Hubbard bands mentioned above. 
%
Our calculations of the 
single-band Hubbard model, 
which are benchmarked against the 
Dynamical Mean Field Theory
(DMFT)~\cite{DMFT,dmft_book}
solution, 
show that the new wave function quantitatively captures not only the dispersion of the QP but also the Hubbard bands.
Furthermore, our theory
enables us to describe 
the Mott transition and the coexistence region between the metallic and the Mott-insulator phases.
%

%

\emph{Ghost GA theory:}--- For 
simplicity, our theory
will be formulated here for the single-band Hubbard model
\bea
\h\!&=&\!
\sum_{RR'}\sum_{\sigma} t_{RR'}\,
\cc_{R\sigma}\ca_{R'\sigma}+
\sum_{R\sigma} {U}\,
\hat{n}_{R\uparrow}\hat{n}_{R\downarrow}
\label{original}
\eea
at half-filling.
The generalization to arbitrary multi-orbital Hubbard Hamiltonians
is straightforward~\cite{supplemental_material}.

\begin{figure}
	\begin{center}
		\includegraphics[width=8.4cm]{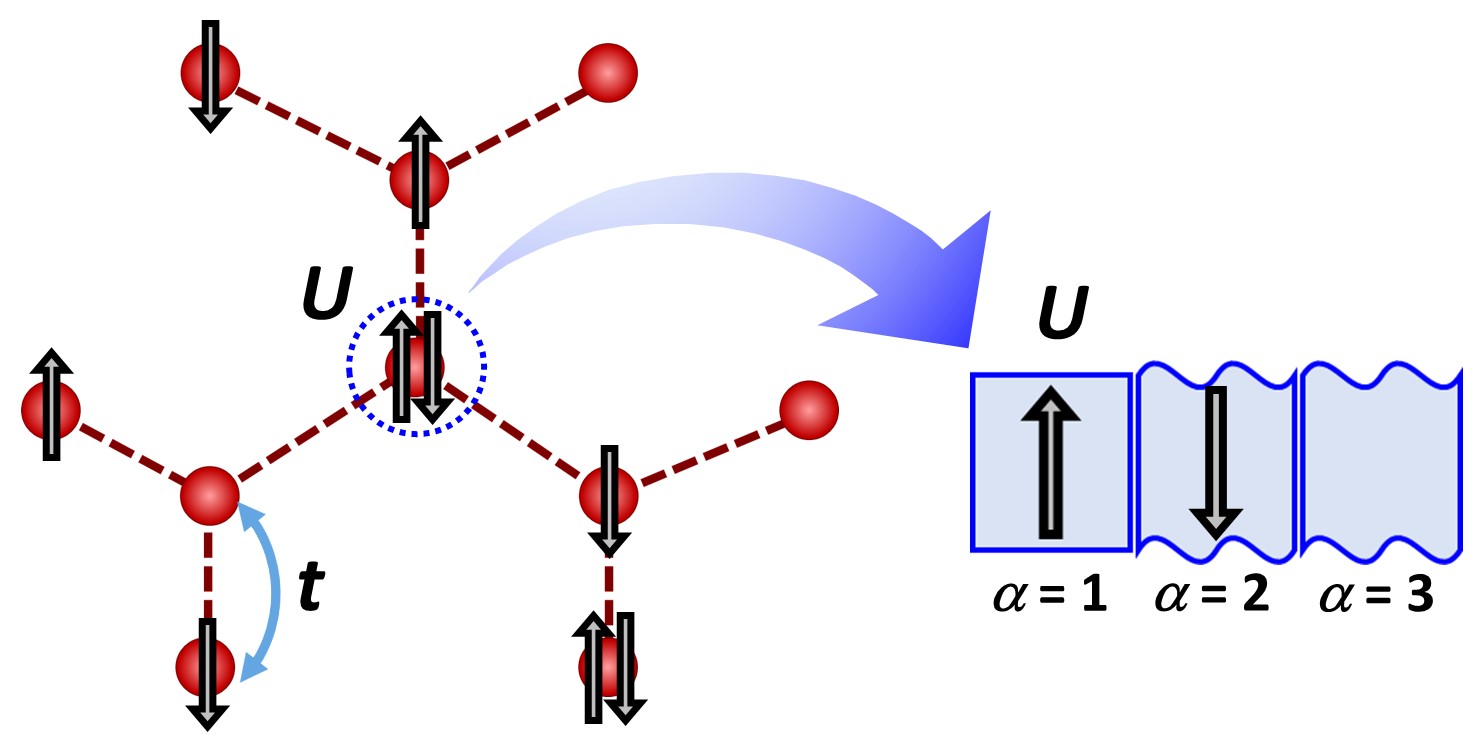}
		\caption{(Color online) Representation of a
			lattice including 2 ghost orbitals ($\alpha=2,3$).
			The Hamiltonian of the system acts as $0$ over the the
			auxiliary ghost degrees of freedom. In particular, the
			Hubbard interaction $U$ acts only over the physical orbital
			$\alpha=1$.
		}
		\label{figure1}
	\end{center}
\end{figure}

In order to construct the Ghost-GA theory we are going to embed the 
physical Hamiltonian of the system [Eq.~\eqref{original}]
within an extended Hilbert space obtained
by introducing auxiliary Fermionic "ghost" degrees of freedom
\emph{not} coupled with the physical orbitals,
see Fig.~\ref{figure1}. 
Let us represent $\h$ within the
extended Hilbert space mentioned above as follows:
\bea
\h\!\!&=&\!\!
 \sum_{RR'}\sum_{\alpha\beta\sigma} \tilde{t}_{RR'}^{\alpha
 \beta}\,
\cc_{R\alpha\sigma}\ca_{R'\beta\sigma}+
\sum_{R}{U}\,
\hat{n}_{R 1\uparrow}\hat{n}_{R 1\downarrow}
\nonumber\\
\!&=&\!\sum_{k}\sum_{\alpha\beta\sigma} \tilde{\epsilon}_{k}^{\alpha\beta}\,
\cc_{k\alpha\sigma}\ca_{k\beta\sigma}
+\sum_{R}{U}\,
\hat{n}_{R 1\uparrow}\hat{n}_{R 1\downarrow}
\,,
\label{hubb}
\eea
where $\tilde{t}_{RR'}^{11}=t_{RR'}$ are the physical
hopping parameters,
$\tilde{\epsilon}_{k}^{11}=\epsilon_k$ are the eigenvalues
of the first term of $\h$,
$\tilde{t}_{RR'}^{\alpha\beta}=\tilde{\epsilon}_{k}^{\alpha\beta}=0$ 
$\forall\, (\alpha,\beta) \neq (1,1)$
and $\sigma$ is the spin.

Our theory consists in applying the ordinary multi-orbital GA 
theory~\cite{lanata,Gebhard,Our-PRX,Attaccalite,Fang}
to Eq~\eqref{hubb}.
In other words, 
the expectation value of $\h$ 
is optimized variationally
with respect to a Gutzwiller wave function represented as
$\ket{\Psi_G}=\proj\ket{\Psi_0}$,
where $\ket{\Psi_0}$ is the most general Slater determinant, $\proj=\prod_R\projR$, and $\projR$ acts over all of the
local degrees of freedom labeled by $R$ --- including the ghost
orbitals $\alpha>1$.
The variational wave function is restricted by the following
conditions:
\bea
\Av{\Psi_0}{\projR^\dagger\projR^\dagga}\!&=&\!\langle\Psi_0|\Psi_0\rangle\\
\Av{\Psi_0}{\projR^\dagger\projR^\dagga\,\cc_{R\alpha\sigma}\ca_{R\beta\sigma}}\!&=&\!
\Av{\Psi_0}{\cc_{R\alpha\sigma}\ca_{R\beta\sigma}}\,,
\eea
which are commonly called "Gutzwiller constraints".
Furthermore, 
the so called "Gutzwiller Approximation"~\cite{Gutzwiller3}
--- which is exact in the limit of infinite dimensions (where DMFT is exact) ---
is employed.
%
The minimization of the variational energy will be performed by
employing the algorithms derived in Ref.~\cite{Our-SB}.

The basis of our theory is that
extending the Hilbert space by
introducing the ghost orbitals does not affect 
the physical Hubbard
Hamiltonian $\h$, 
as all of its terms 
involving ghost orbitals
are multiplied by $0$, see Eq.~\eqref{hubb}.
The advantage of enlarging the Hilbert space arises exclusively from
the fact that the 
corresponding Ghost-GA multi-orbital variational space
%
is substantially more rich
with respect to the original GA variational space
(where $\projR$ acts only on the physical degrees of
freedom)~\cite{supplemental_material}.
In this respect, our scheme presents analogies, e.g., with the
solution of the Affleck, Lieb, Kennedy and Tasak (AKLT) model~\cite{AKLT}
and with 
the theories of Matrix Product States (MPS) and
Projected Entangled Pair States (PEPS)~\cite{Vaestrate_review,Ulrich_review},
which are also variational
constructions 
involving \emph{virtual}
entanglement and local maps~\cite{footnote-3}.
Further technical details about the role of the ghost
orbitals are provided in the supplemental
material~\cite{supplemental_material}.
%
%

As shown in previous works, see, e.g., Refs.~\cite{Our-PRX}, 
the variational energy minimum of $\h$ is realized by a wave function 
$\ket{\Psi_G}=\proj\ket{\Psi_0}$ where 
$\ket{\Psi_0}$ is the
ground state of a quadratic multi band
Hamiltonian represented as 
\bea
\Hqp\!\!&=&
\sum_k\sum_{ab\sigma}
\big[\tR \tilde{\epsilon}_{k}
\tR^\dagger+\tlambda\big]_{ab}\,
\fc_{k a\sigma}\fa_{k b\sigma}
\nonumber\\
&=&\sum_{kn\sigma}\tilde{\epsilon}^*_{kn}\,
\psi^\dagger_{kn\sigma}\psi^\dagga_{kn\sigma}
\,,
\label{hqp}
\eea
where $f_{ka\sigma}$ are related to $c_{ka\sigma}$ by
a proper unitary transformation~\cite{lanata,Our-PRX},
the matrices $\tR$ and $\tlambda$
are determined variationally
and $\tilde{\epsilon}^*_{kn}$ are the eigenvalues of $\Hqp$.
The states 
$
\ket{\Psi^p_{Gkn\sigma}}=\proj\psi^\dagger_{kn\sigma}\ket{\Psi_0}$
and
$\ket{\Psi^h_{Gkn\sigma}}=\proj\psi^\dagga_{kn\sigma}\ket{\Psi_0},
$
where $\psi^\dagger_{ka\sigma}$ are the eigen-operators of $\Hqp$, 
represent excited states of 
$\h$~\cite{Gebhard-FL,mythesis,supplemental_material}.
%

The energy-resolved Green's function of the physical degrees of 
freedom ($\alpha=1$) can be evaluated in terms of the
excitations mentioned above~\cite{supplemental_material} and represented as
\be
G(\epsilon_{k},\omega)\!=\!
\left[\tR^\dagger
\frac{1}{\omega\!-\!\big(\tR \tilde{\epsilon}_{k} \tR^\dagger \!+\! \tlambda\big)}
\tR\right]_{11}
\!\!\!\!=\!\big[\omega\!-\!\epsilon_{k}\!-\!\Sigma(\omega)\big]^{-1}
\!\!,
\label{g}
\ee
where 
the subscript "$11$" indicates that we are interested only in the
physical component $\alpha=\beta=1$ of the Green's function. 
We point out that, since Eq.~\eqref{g} involves a matrix
inversion~\cite{poles-Gabi,poles-Gabi-2},
the Ghost-GA approximation $\Sigma(\omega)$ to the physical self-energy
is generally a \emph{non-linear} function~\cite{supplemental_material} ---
while it is linear by construction
in the ordinary GA theory.
Note also that 
the poles of $G(\epsilon_{k},\omega)$
coincide with the eigenvalues $\tilde{\epsilon}_{kn}^*$ of $\h_{\text{qp}}$,
see Eq.~\eqref{hqp}.


\emph{Application to the single-band Hubbard model:}---
Below we apply our approach to the Hubbard Hamiltonian [Eq.~\eqref{hubb}] at half-filling
assuming a semicircular DOS~\cite{footnote-1},
which corresponds, e.g., to the Bethe lattice in the limit of infinite connectivity, where DMFT is exact~\cite{DMFT}. 
The half-bandwidth $D$ will be used as the unit of energy.
The extended Ghost-GA scheme will be applied
following the procedure of Ref.~\cite{Our-SB},
utilizing up to 2 ghost orbitals.

\begin{figure}
	\begin{center}
		\includegraphics[width=8.4cm]{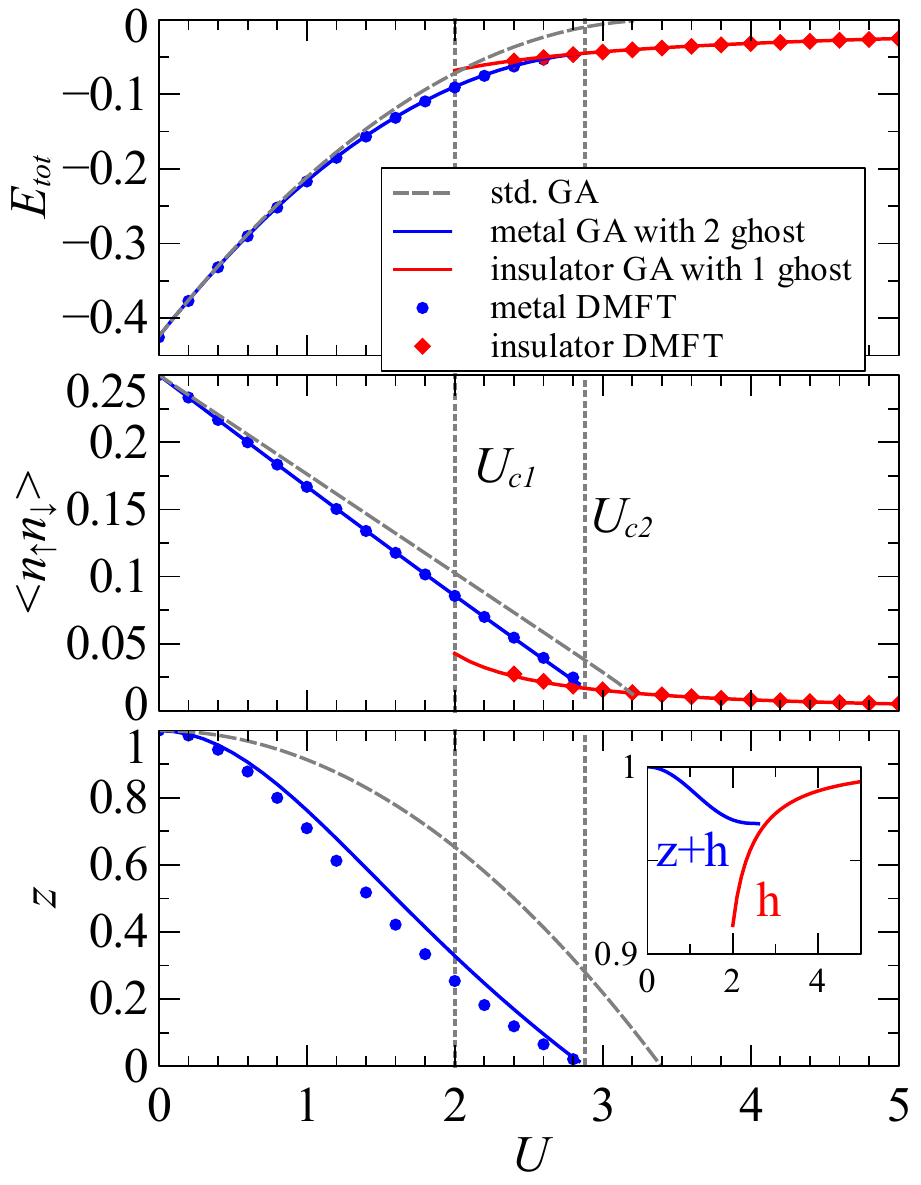}
		\caption{(Color online) Evolution of 
		total energy (upper panel), local double occupancy
		(middle panel) and QP weight (lower panel) as a function of
		the Hubbard interaction strength $U$
		for the single-band Hubbard model
		with semicircular DOS at half-filling. 
		The Ghost-GA results are shown in comparison with
		the ordinary GA and with the DMFT+NRG results.
		The Ghost-GA boundaries of the coexistence region $U_{c1},U_{c2}$ are indicated by vertical dotted lines.
		Inset: Integral of Ghost-GA local spectral weight over all frequencies (see discussion in main text).
		}
		\label{figure2}
	\end{center}
\end{figure}

In Fig.~\ref{figure2} is shown the evolution
as a function of the Hubbard interaction
strength $U$ of the Ghost-GA total energy,
the local double occupancy and the QP weight $z$.
Our results are shown
in comparison with the ordinary GA theory and with DMFT
in combination with Numerical Renormalization Group
(NRG). 
In particular, we employed 
the "NRG Ljubljana" impurity solver~\cite{zitko-sneg}.

The agreement between Ghost-GA and DMFT
is quantitatively remarkable.
In particular, the Ghost-GA theory enables us to account for
the coexistence region of the Mott and metallic phases,
which is not captured by the ordinary GA theory.
The 
values of the boundaries of the coexistence region
$U_{c1}\simeq 2$, $U_{c2}\simeq 2.88$ are in good
agreement with the DMFT results available in the
literature~\cite{Uc-comp-dmrg,Uc-comp-ctqmc,Uc-comp-nrg,Uc-comp-ed},
i.e., $U_{c1}\simeq 2.39$, $U_{c2}\simeq 2.94$.
The Ghost-GA value of $U_{c2}$, which is the
actual Mott transition point at $T=0$, is particularly accurate.
The method also gives a reasonable value for the very small energy scale
characterizing the coexistence region, which we can estimate as
$T_c \simeq E_{\text{ins}}(U_{c1}) - E_{\text{met}}(U_{c1}) \simeq 0.02$,
consistently with both DMFT and experiments~\cite{Tc-vlad,Tc-organic-exp}.
We point out also that, as shown in the second panel
of Fig.~\ref{figure2}, the Ghost-GA approach captures
the charge fluctuations in the Mott phase,
while this is approximated by the 
simple atomic limit (which has zero double occupancy)
within the Brinkman-Rice scenario~\cite{brinkman&rice}.

Interestingly, while at least 2 ghost orbitals are necessary 
to obtain
the data illustrated above for the Metallic solution,
1 ghost orbital is sufficient 
to obtain our results concerning the Mott phase.
Increasing further the number of ghost orbitals
does not lead to any appreciable difference~\cite{supplemental_material}.
%
As we are going to show, this 
is connected with
the fact that the electronic structures of the Mott
and the metallic phases are topologically 
distinct. 

\begin{figure}
	\begin{center}
		\includegraphics[width=8.4cm]{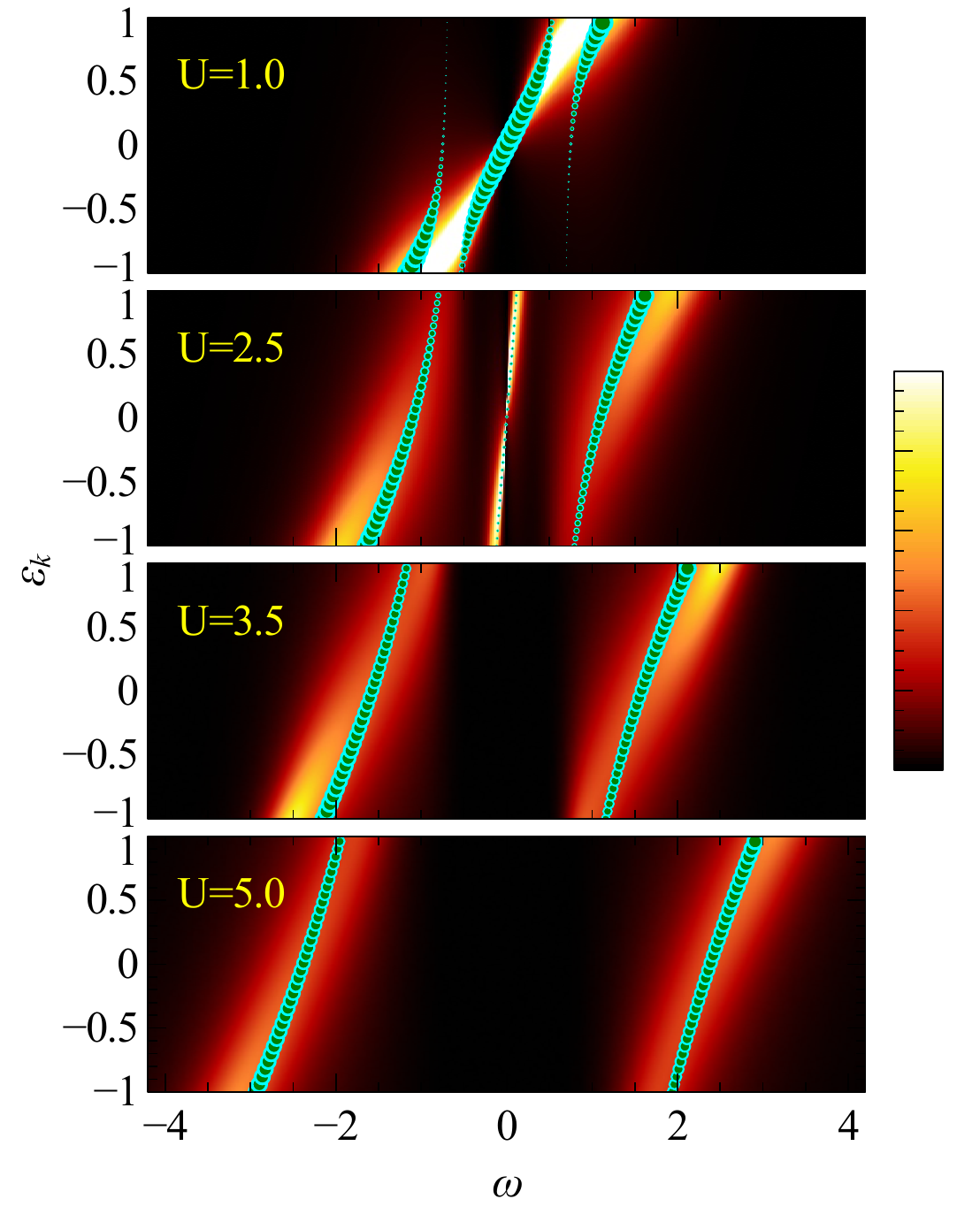}
		\caption{(Color online) 
			Poles of the Ghost-GA
			energy-resolved Green's function (bullets),
			see Eq.~\eqref{g},
			in comparison with DMFT+NRG.
			The size of the bullets indicates the spectral weights
			of the corresponding poles.
			Metallic solution for $U=1,\,2.5$ and Mott solution for $U=3.5,\,5$.
		}
		\label{figure3}
	\end{center}
\end{figure}

Let us now analyze the Ghost-GA single-particle
Green's function $G(\epsilon,\omega)$, see Eq.~\eqref{g}. 
In Fig.~\ref{figure3} is shown the Ghost-GA energy-resolved
spectral function
$A(\epsilon,\omega)=-\frac{1}{\pi}\text{Im}{G(\epsilon,\omega)}$
in comparison with DMFT~\cite{DMRG-comp-A}.
Although the broadening of the bands (scattering rate), is
not captured by our approximation
(as it is not captured by the ordinary GA),
the positions and the weights of the poles of the Ghost-GA
spectral function encode most of the DMFT features,
not only at low energies (QP excitations), but also at high energies (Hubbard bands).
In order to analyze how the
spectral properties of the system 
emerge within the Ghost-GA theory,
it is particularly convenient to
express the QP Hamiltonian [Eq.~\eqref{hqp}]
in a gauge where $\tlambda$ is
diagonal~\cite{footnote-2}.

In the metallic phase,
an explicit Ghost-GA calculation obtained employing 2 ghost orbitals
shows that the matrices $\tR$ and 
$\tlambda$ are represented as follows:
\bea
\tlambda_{ij}&=&l\,\delta_{ij}(\delta_{2i}-\delta_{3i})
\label{lam-met}
\\
\tR_{ij}&=&\delta_{j1}\left(\sqrt{z}\,\delta_{i1}+
\sqrt{h}\,
(\delta_{i2}+\delta_{i3})/\sqrt{2}
%
\right)\,,
\label{R-met}
\eea
where $\delta_{ij}$ is the Kronecker delta, and
$l$, $z$ and $h$ 
are real positive numbers 
determined numerically as in Ref.~\cite{Our-SB}.
The corresponding self-energy, see Eq.~\eqref{g},
is~\cite{footnote-4}:
\be
\Sigma(\omega)=\frac{\omega}{1+\frac{1}{z-
		\frac{\omega^2-l^2+2h\omega^2}{\omega^2-l^2}}
}
= -\frac{1-z}{z}\,\omega + o(\omega^2)
\,.
\label{sigM}
\ee
Thus, the variational parameter $z$ of Eq.~\eqref{R-met}
represents the QP weight, whose behavior was displayed in
the third panel of Fig.~\ref{figure2}. 
Note that the overall spectral weight 
$\int d\omega\int d\epsilon\,\rho(\epsilon)\,A(\epsilon,\omega)$,
where $\rho(\epsilon)$ is the semicircular DOS, 
is not $z$ as in the
ordinary GA theory, but it is $z+h=[\tR^\dagger\tR]_{11}$,
which is almost equal to $1$ for all values of $U$ (see the inset
of the third panel in Fig.~\ref{figure2}).
The additional spectral contribution $h$, which is not present
in the ordinary GA approximation, 
enables the Ghost-GA theory
to account for the Hubbard bands.

In the Mott phase, an explicit Ghost-GA calculation obtained
employing 1 ghost orbital shows that the matrices $\tR$ and 
$\tlambda$ are represented as follows:
\bea
\tlambda_{ij}&=&l\,\delta_{ij}(\delta_{1i}-\delta_{2i})
\label{lam-ins}
\\
\tR_{ij}&=&\delta_{j1}\sqrt{h}\,
(\delta_{i1}+\delta_{i2})/\sqrt{2}
%
\,,
\label{R-ins}
\eea
where 
$l$ and $h$ 
are real positive numbers 
determined numerically as in Ref.~\onlinecite{Our-SB}.
Note that $h=[\tR^\dagger\tR]_{11}\simeq 1$
(see the inset of the third panel in Fig.~\ref{figure2}).
The corresponding self-energy, see Eq.~\eqref{g}, 
is~\cite{footnote-4}:
\be
\Sigma(\omega)=-\frac{1-h}{h}\,\omega + \frac{l^2}{h}\,\frac{1}{\omega}
\,.
\label{sigI}
\ee
The pole of the self energy at $\omega=0$, which is the source of the Mott gap, is captured by the Ghost-GA
theory.

The analysis above clarifies also why, by construction,
within the Ghost-GA approximation
the self-energy can develop poles, see Eqs.~\eqref{sigM} and \eqref{sigI},
but can not capture branch-cut singularities on the real axis.
%

It is important to point out that the Hilbert space extension that we
have introduced in this work has been essential in order to capture
the effect of the electron correlations on the topology of the excitations
--- such as the change of the number of bands at the Mott transition
(between 3 bands in the metallic phase and 2 bands in the Mott phase).
In fact, without extending the Hilbert space, the ordinary GA theory
enables only to renormalize and shift the band structure with respect 
to the uncorrelated limit $U=0$, without affecting its qualitative topological structure.
On the other hand, extending the Hilbert space enables us to relax this
constraint, as $G(\epsilon,\omega)$, see Eq.~\eqref{g}, 
is variationally allowed to have any number of distinct poles
equal or smaller
to the corresponding total (physical and ghost) number of
orbitals~\cite{supplemental_material}.
It is for this reasons that only 1 ghost (2 orbitals) is sufficient to describe the Mott phase of the single-band Hubbard model, 
while at least 2 ghosts (3 orbitals) are necessary
in order to describe its metallic phase --- whose spectra includes the QP excitations and the 2 Hubbard bands.
A remarkable aspect of this construction is that,
within the Ghost-GA theory,
the information concerning the spectral
function --- including the Hubbard bands --- is entirely encoded
in only 3 parameters ($z,h,l$) in the Metallic phase,
and in 2 parameters ($h,l$) in the Mott phase,
see Eqs.~\eqref{lam-met}, \eqref{R-met}, \eqref{lam-ins}, \eqref{R-ins}.

\emph{Conclusions:}---
We derived 
a unified theoretical picture for excitations in Mott systems
based on a generalization of the GA,
which captures not only the low-energy QP excitations,
but also the Hubbard bands.
The essential idea consists in extending the Hilbert space of the
system by introducing auxiliary "ghost" orbitals. 
This construction enables us to express analytically 
many important features of both types of excitations
in terms of the bare DOS of the uncorrelated system and a few renormalization
parameters, in a similar fashion as for the QP excitations
in Landau theory of Fermi liquids.
In particular, this idea provides
us with
a conceptual picture
which assigns naturally a
Bloch character to the Hubbard bands even in Mott insulators.
In this respect, we note that 
our theory presents a few suggestive analogies
with the interesting (but rather speculative) idea
of "hidden Fermi liquid" previously introduced by P. W. Anderson~\cite{Hidden-FL}
within the context of the BCS wave function (for superconductors)
and the Laughlin's Jastrow wave function (for the Fractional
Hall Effect).
In fact,
they both propose a descriptions of non-Fermi liquid states
related to ordinary Fermi liquids
residing in unphysical Hilbert spaces,
see, e.g., Ref.~\cite{And-HFL-PNAS}.
From the computational perspective,
our Ghost-GA theory constitutes a very promising tool
for ab-initio calculations in combination with
Density Functional Theory (DFT)~\cite{HohenbergandKohn,KohnandSham,Fang,Ho,Our-PRX},
as it is substantially more accurate with respect to the
ordinary GA approximation, without much additional
computational cost.
In fact, within the numerical
scheme described in Refs.~\cite{Our-SB,Our-PRX}, our theory results
in solving iteratively a finite impurity model, where the number of bath
sites grows linearly with the
total number of ghost orbitals~\cite{supplemental_material}.
%
Since there exist numerous available techniques enabling to solve efficiently
this auxiliary problem,
see, e.g., Refs.~\cite{ulrich,PRB-DMRGvsNRG,Vaestrate_NRG-DMRG-MPS,imp_quanty-guy,GRG},
this opens an exciting avenue for realistic modeling of many challenging materials, including predictions of ARPES spectra for complex orbitally-selective Mott insulators. 
Furthermore, since the Ghost-GA theory 
is based on the multi-orbital GA~\cite{Our-SB,Our-PRX},
it can be straightforwardly generalized to finite temperatures~\cite{Our-Temperature,finiteT-GA-fabrizio,finiteT-GA-Wang},
to non-equilibrium problems~\cite{Mic-Schi_td-ga,my-transport-2},
and to calculate linear response functions~\cite{Fabrizio-fluctuations}.
For the same reason, 
the Ghost-GA theory 
can be also straightforwardly
reformulated~\cite{equivalence_GA-SB,lanata,Our-PRX}
in terms of the rotationally invariant slave boson (RISB) theory~\cite{Georges,rotationally-invariant_SB,supplemental_material},
whose exact operatorial foundation recently derived in Ref.~\cite{Our-SB}
constitutes a starting point to calculate
further corrections~\cite{fluctuations-SB-2017}.
It would be also interesting to apply the ghost-orbital
Hilbert space extension 
in combination with the Variational Monte Carlo
method~\cite{VMC-Fermions} or the generalization
of the GA to finite dimensions of Ref.~\cite{Bunemann-diagrammatic},
which might lead to 
a more accurate description of strongly correlated electron systems
even beyond the DMFT approximation.

\begin{acknowledgments}
  N.L., T.H. and V.D. were partially supported by the NSF grant DMR-1410132 and the National High Magnetic Field Laboratory.
  Y.Y. was supported by the U.S. Department of energy,
  Office of Science, Basic Energy Sciences, as a part of the Computational Materials Science Program.
\end{acknowledgments}



\widetext
\clearpage

\begin{center}
	\textbf{\large Supplemental Material:\\ Emergent Bloch Excitations in Mott Matter}
\end{center}

\renewcommand\thesection{\arabic{section}}
\setcounter{equation}{0}
\setcounter{figure}{0}
\setcounter{table}{0}
\setcounter{page}{1}
\makeatletter

\begin{changemargin}{1.3cm}{1.3cm} 
	~~~In this supplemental material we summarize several
	technical details concerning the Ghost-GA method, including the
	numerical implementation and an alternative reformulation in terms of
	Rotationally Invariant Slave Boson (RISB) theory.
	Furthermore, we discuss the Ghost-GA solution of the Hubbard model
	utilizing up to 3 ghost orbitals (i.e., 2 additional ghost orbitals with respect
	to the main text) in all regimes of interaction strength.
	Finally, we show calculations of the Hubbard model
	away from half-filling and of the single-orbital Anderson Impurity Model.
\end{changemargin}

\section{I.~~~Technical details about Ghost-GA theory}

\subsection{A.~~~Introduction}

Let us consider a generic multi-orbital Hubbard Hamiltonian represented as
\be
\h \,=\, 
\sum_{RR'} 
\sum_{\alpha,\beta=1}^{\tilde{\nu}} {t}_{RR'}^{\alpha\beta}\,
\cc_{R\alpha}\ca_{R'\beta}+
\sum_{R} \h_R^{\text{loc}}
\;=\;
\sum_{k} 
\sum_{\alpha,\beta=1}^{\tilde{\nu}} {\epsilon}_{k}^{\alpha\beta}\,
\cc_{k\alpha}\ca_{k\beta}
+\sum_{R} \h_R^{\text{loc}}
\,,
\label{hubb-supp}
\ee
where $\tilde{\nu}$ is the total number of Fermionic degrees of 
freedom (both spin and orbital), and
$\h_R^{\text{loc}}$ is a generic local operator
(i.e., any operator that can be expressed
in terms of operators $\{\cc_{R\alpha},\ca_{R\beta}\}$ with fixed
label $R$).

The multi-orbital Gutzwiller Approximation~[\onlinecite{Gutzwiller3-supp, lanata-supp,Gebhard-supp,Our-PRX-supp,Attaccalite-supp}] (GA)
consists in minimizing the expectation value of $\h$
with respect the most general Gutzwiller wave function:
\be
\ket{\Psi_G}=\proj\ket{\Psi_0}\,,
\label{GWF-supp}
\ee
where $\ket{\Psi_0}$ is the most general Slater determinant, $\proj=\prod_R\projR$, and $\projR$ is the most general map acting
over all of the local degrees of freedom labeled by $R$.
As discussed in Ref.~[\onlinecite{Our-PRX-supp}], it is convenient to represent
$\projR$ in a "mixed basis" representation as follows:
\be
\projR = \sum_{A,n=0}^{2^{\tilde{\nu}}-1} \tilde{\Lambda}_{A n}\,
\ket{A,R}\bra{n,R}
\label{Lam-supp}
\,,
\ee
where 
\bea
\ket{A,R}&=&[\cc_{R1}]^{q_1(A)}\,...\;[\cc_{R\tilde{\nu}}]^{q_{\tilde{\nu}}(A)}\,\ket{0}\,, \;\,\; A\in\{0,...,2^{\tilde{\nu}}-1\} 
\\
\bra{n,R}&=&\bra{0}\,
[\fa_{R\tilde{\nu}}]^{q_{\tilde{\nu}}(n)}
\,...\;
[\fa_{R1}]^{q_1(n)}
\,, \;\,\; n\in\{0,...,2^{\tilde{\nu}}-1\} \,,
\eea
the occupation numbers $q_1(l),...,q_{\tilde{\nu}}(l)$ are the
digits of a generic integer $l\in\{0,...,2^{\tilde{\nu}}-1\}$
in basis $2$, and the ladder operators 
$\fc_{Ra}$ are related to $\cc_{R\alpha}$ through a unitary transformation
such that
\be
\Av{\Psi_0}{\fc_{Ra}\fa_{Rb}}\propto\delta_{ab}\quad\forall\,a,b \in
\{1,...,\tilde{\nu}\}\,.
\label{natural-supp}
\ee

The variational wave function is restricted by the following
conditions:
\bea
\Av{\Psi_0}{\projR^\dagger\projR^\dagga}\!&=&\!\langle\Psi_0|\Psi_0\rangle\\
\Av{\Psi_0}{\projR^\dagger\projR^\dagga\,\fc_{R\alpha}\fa_{R\beta}}\!&=&\!
\Av{\Psi_0}{\fc_{R\alpha}\fa_{R\beta}}\qquad\forall\,\alpha,\beta=1,...,\tilde{\nu} 
\,,
\label{gconstr-supp}
\eea
which are commonly called "Gutzwiller constraints".
Furthermore, the "Gutzwiller Approximation", which becomes
exact in the limit of infinite dimension~[\onlinecite{Gutzwiller3-supp}]
--- where Dynamical Mean Field Theory (DMFT) is exact~[\onlinecite{DMFT-supp}]
--- is assumed.

As shown in previous works, see, e.g., Refs.~[\onlinecite{lanata-supp,GRG-supp}],
the accuracy of the Gutzwiller method
is improved substantially when the
operator $\proj$ (Gutzwiller projector) is allowed to act also beyond
the space generated by the
correlated degrees of freedom.
The main technical idea underlying the Ghost-GA theory presented 
in this work consists essentially in creating new auxiliary (inert)
orbitals with the sole purpose of enriching the variational space
without changing its formal structure (i.e., a Gutzwiller-projected
Slater determinant).
Of course, 
as the ordinary GA, the Ghost-GA theory
exploits the so called "Gutzwiller Approximation"~[\onlinecite{Gutzwiller3-supp}],
which becomes
exact in the limit of infinite dimension (where DMFT is exact).
Thus, in summary, the Ghost-GA theory consists in an improved variational
approximation to DMFT realized by enlarging the ordinary GA
variational space.

Further details about the role of the ghost orbitals are provided
in the subsections below.

\subsection{B.~~~Functional formulation of the Gutzwiller energy minimization}\label{covariance-sec}

As discussed in the main text, all of our calculations of the single
band Hubbard Hamiltonian [see Eq.~1 of the main text]
have been performed by applying the ordinary
multi-orbital Gutzwiller theory
to the same Hamiltonian "embedded" within the extended
Hilbert space obtained by introducing auxiliary ghost orbitals
[see Eq.~2 of the main text].
Specifically, we have employed the numerical procedure 
derived in described in Refs.~[\onlinecite{Our-SB-supp,Our-PRX-supp}],
which we summarize below for completeness.
%

For simplicity, the theory will be formulated here assuming a
translationally invariant Gutzwiller solution.
As shown in Refs.~[\onlinecite{Our-SB-supp,Our-PRX-supp}], the minimization of the 
expectation value of a generic multi-orbital Hamiltonian represented
as in Eq.~\eqref{hubb-supp}
with respect to the Gutzwiller wave function can be conveniently
formulated in terms of the following Lagrange function:
\bea
&&\Lag[\Psi_0,E;\,  \Phi,E^c;\,  \R,\R^\dagger,\lambda;\, \D,\D^\dagger, \lambda^{c};\,\Delta]
=\frac{1}{\mathcal{N}}\Av{\Psi_0}{\h_{\text{qp}}[\R,\R^\dagger;\lambda]}
+E\!\left(1\!-\!\langle\Psi_0|\Psi_0\rangle\right)
\nonumber\\
\nonumber\\&&\quad\quad
+\left[\Av{\Phi}{\h_{\text{emb}}[\D,\D^\dagger;\lambda^c]}
+E^c\!\left(1-\langle \Phi | \Phi \rangle
\right)\right]
\nonumber\\&&\quad\quad
- 
\left[
\sum_{a,b=1}^{\tilde{\nu}}\left(
\lambda_{ab}+\lambda^c_{ab}\right)\Delta_{ab}
+\sum_{c, a,\alpha=1}^{\tilde{\nu}} \left(
\D_{a\alpha}\R_{c\alpha}
\left[\Delta(1-\Delta)\right]^{\frac{1}{2}}_{ca}
+\text{c.c.}\right)
\right]\,,
\label{Lag-SB-emb-supp}
\eea
where $\mathcal{N}$ is the total number of $k$ points,
$E$ and $E^c$ are real numbers, $\Delta$, $\lambda^c$ and $\lambda$ are
Hermitian matrices, $\D$ and $\R$ are generic complex matrices.
The auxiliary Hamiltonians $\Hqp$ and $\h_{\text{emb}}$,
which are called "quasiparticle Hamiltonian" and "Embedding Hamiltonian",
respectively, are defined as follows:
\bea
\Hqp&=&
\sum_{k}\sum_{a,b=1}^{\tilde{\nu}}
\big[\R {\epsilon}_{k}
\R^\dagger+\lambda\big]_{ab}\,\fc_{k a}\fa_{k b}
= 
\sum_{k}\sum_{n=1}^{\tilde{\nu}} {\epsilon}^*_{kn}\,
\psi^\dagger_{kn}\psi^\dagga_{kn}
\label{hqp-supp}
\\
\h_{\text{emb}} 
&=&
\h^{\text{loc}}[\{\hat{c}^\dagger_{\alpha}\},\{\hat{c}^\dagga_{\alpha}\}] +
\sum_{a,\alpha=1}^{\tilde{\nu}} \left(
\D_{a\alpha}\,
\hat{c}^\dagger_{\alpha}\hat{f}^\dagga_{a}+\text{H.c.}\right)
+ 
\sum_{a,b=1}^{\tilde{\nu}} \lambda^c_{ab}\,
\hat{f}^\dagga_{b}\hat{f}^\dagger_{a}\label{h-emb-i}
\,.
\label{hemb-supp}
\eea
%
The auxiliary Fermionic Hamiltonian $\h_{\text{emb}}$, which
was introduced
in Ref.~[\onlinecite{Our-PRX-supp}], enables us to exploit techniques such as those
developed in quantum chemistry in order to tackle the energy minimization
problem.
The state $\ket{\Phi}$, see Eq.~\eqref{Lag-SB-emb-supp},
is the most general many-body state --- with $\tilde{\nu}$
electrons --- belonging to 
the auxiliary embedding Hilbert space mentioned above.

\subsection{C.~~~Connection between embedding state $\ket{\Phi}$ and Gutzwiller variational parameters}

Let us consider the following expansion of the most general state $\ket{\Phi}$:
\be
\ket{\Phi} \equiv \sum_{A, n=0}^{2^{\tilde{\nu}}-1} e^{i\frac{\pi}{2}N_n(N_n-1)}
\tilde{\phi}_{A n}\,
[{\hcc}_{1}]^{q_1(A)}\,...\;[{\hcc}_{\tilde{\nu}}]^{q_{\tilde{\nu}}(A)}\,
[{\hfa}_{1}]^{q_1(n)}\,...\;[{\hfa}_{\tilde{\nu}}]^{q_{\tilde{\nu}}(n)}\,\ket{\bar{0}} 
\,,\label{pure-emb-supp}
\ee
where $A,n \in\{0,...,2^{\tilde{\nu}}-1\}$, $N_n=\sum_{a=1}^{\tilde{\nu}} q_a(n)$ and $\ket{\bar{0}}=[{\hfc}_{1}]\,...\;[{\hfc}_{\tilde{\nu}}]\,\ket{0}$.

For later convenience, it is important to point out that, as shown in
Ref.~[\onlinecite{Our-PRX-supp}], the matrix $\tilde{\phi}$ with entries $\tilde{\phi}_{A n}$
is connected with 
the Gutzwiller variational parameters as follows:
\be
\tilde{\phi}=\tilde{\Lambda}\sqrt{P^0}\,,
\label{pl-supp}
\ee
where $\tilde{\Lambda}$ is the matrix defining $\projR$ as in Eq.~\eqref{Lam-supp},
$P^0$ is $2^{\tilde{\nu}}\times 2^{\tilde{\nu}}$ matrix with entries
\be
P^0_{nm}=\Av{\Psi_0}{[\fa_{R\tilde{\nu}}]^{q_{\tilde{\nu}}(n)}
	\,...\;
	[\fa_{R1}]^{q_1(n)}\,
	[\fc_{R1}]^{q_1(m)}
	\,...\;
	[\fc_{R\tilde{\nu}}]^{q_{\tilde{\nu}}(m)}}\,,
\ee 
and $n,m \in\{0,...,2^{\tilde{\nu}}-1\}$.
Note that, because of Eq.~\eqref{natural-supp}, $P^0$ is diagonal.


%

\subsection{D.~~~Gutzwiller Lagrange equations in the standard multi-orbital GA}

%


Following Ref.~[\onlinecite{Our-PRX-supp}], in order to take into account 
the fact that $\Delta$, $\lambda^c$ and $\lambda$ are
Hermitian matrices we introduce the following parametrizations:
\bea
\Delta&=&\sum_{s=1}^{\tilde{\nu}^2} d_s\,^th_{s}
\\
\lambda^c&=&\sum_{s=1}^{\tilde{\nu}^2} l_{s}^c\,h_{s}
\label{lambdac-lc-supp}
\\
\lambda&=&\sum_{s=1}^{\tilde{\nu}^2} l_{s}\,h_{s}
\label{lambda-l-supp}
\\
\R&=&\sum_{s=1}^{\tilde{\nu}^2} r_{s}\,h_{s}
\label{R-r-supp}
\eea
where the set of matrices $h_{s}$ is an orthonormal
basis of the space of Hermitian matrices with
dimension $\nu$, $^t h_{s}$ are the corresponding transposed matrices,
and $d_{s}$, $l_{s}^c$ and $l_{s}$ are real numbers, while
$r_{s}$ are complex numbers.
The above-mentioned orthonormality is defined
with respect to the standard scalar product
$(A,B)\equiv\Tr\!\left[A^\dagger B\right]$.
%


It can be readily shown that the saddle-point conditions of $\Lag$,
see Eq.~\eqref{Lag-SB-emb-supp},
with respect to  all of its arguments provides the following system
of Lagrange equations:
\bea
&&\frac{1}{\mathcal{N}} 
\sum_k \big[f\!\left(\R\epsilon_{k}\R^\dagger+\lambda\right)\big]_{ba}
= \Delta_{ab}\label{l2-supp}\\
&&\frac{1}{\mathcal{N}}\,
\sum_k \big[ 
\epsilon_k\R^\dagger\,
f\!\left(\R\epsilon_k\R^\dagger+\lambda\right)\big]_{\alpha a}
=\sum_c \D_{c\alpha}
\left[\Delta\left(1-\Delta\right)\right]^{\frac{1}{2}}_{ac}
\label{l3-supp}\\
&& \sum_{cb\alpha}\frac{\partial}{\partial d^p_{s}}\left[\Delta\left(1-\Delta\right)\right]^{\frac{1}{2}}_{cb} \D_{b\alpha}\R_{c\alpha}+\text{c.c.}
+\left[l+l^c\right]_{s}=0
\label{l4-supp}\\
&&\h^{\text{emb}}[\D,\lambda^c]\,\ket{\Phi}
= E^c\,\ket{\Phi}
\label{l5-supp}\\
&&\big[\mathcal{F}^{(1)}\big]_{\alpha a}\equiv
\Av{\Phi}{\hat{c}^\dagger_{\alpha}\hat{f}^\dagga_{a}}
-\sum_c\left[\Delta\left(1-\Delta\right)\right]^{\frac{1}{2}}_{ca} [\R]_{c\alpha}
=0\label{l6-supp}\\
&&\big[\mathcal{F}^{(2)}\big]_{ab}\equiv
\Av{\Phi}{\hat{f}^\dagga_{b}\hat{f}^\dagger_{a}}
- \left[\Delta\right]_{ab}
=0 \label{l7-supp}\,,
\eea
where the function $f$ appearing in Eqs.~\eqref{l2-supp} and \eqref{l3-supp}
is the Fermi function at zero temperature ($T=0$).

The solution of the equations above can be obtained numerically
employing the following procedure~[\onlinecite{Our-PRX-supp}].
(I) Given a set of coefficients $r_{s}$ and $l_{s}$, determine the
corresponding matrices $\R$ and $\lambda$ using
Eqs.~\eqref{lambda-l-supp} and \eqref{R-r-supp}, and calculate
$\Delta$ using Eq.~\eqref{l2-supp}.
(II) Calculate $\D$ by inverting Eq.~\eqref{l3-supp}.
(III) Calculate the coefficients $l^c_{s}$ using Eq.~\eqref{l4-supp}
and the corresponding matrix $\lambda^c$ using Eq.~\eqref{lambdac-lc-supp}.
(IV) Construct the embedding Hamiltonian $\h^{\text{emb}}$
and compute its ground state $\ket{\Phi}$, see Eq.~\eqref{l5-supp},
within the subspace with $\nu$ electrons.
(V) Determine the left members of Eqs.~\eqref{l6-supp} and \eqref{l7-supp}.
The equations \eqref{l6-supp} and \eqref{l7-supp} are satisfied
if and only if the coefficients $r_{s}$ and $l_{s}$ proposed
at the first of the steps above
identify a solution of the GA Lagrange function.
The steps above enable us to reduce the solution of the
GA Lagrange equations to a root problem,
which can be formally represented as
\be
(\mathcal{F}^{(1)}(r,l), \mathcal{F}^{(2)}(r,l))=0\,,
\label{roothubb-supp}
\ee
that can be readily
solved numerically, e.g., using the quasi-Newton method.

All of the expectation values with respect to the Gutzwiller wavefunction
can be readily expressed in terms of the states $\ket{\Phi}$
and $\ket{\Psi_0}$ obtained after convergence.

\subsection{E.~~~Proof that the solution of the Ghost-GA equations is disentangled
	from the auxiliary ghost subsystem}\label{dis-subsec}

As explained in the main text, the Ghost-GA 
theory is formulated by embedding the 
Hamiltonian $\h$ within an extended Hilbert space, which includes
also auxiliary (ghost) Fermionic degrees of freedom.
Such a procedure is obviously licit, as the inequality underlying the
variational principle, i.e.,
\be
\Av{\Psi}{\hat{H}}\geq E_0\,,\label{vp-supp}
\ee
where $E_0$ is the ground state energy of $\hat{H}$, remains valid even
if $\ket{\Psi}$ is a generic state of the \emph{extended} Hilbert space.

Since $\hat{H}$ lives only within the physical subsystem,
we already know a-priori that the exact ground state
(which realizes the above variational minimum)
has to be \emph{disentangled} from the auxiliary
degrees of freedom.
Here we demonstrate that, indeed,
this condition 
is {\emph{satisfied exactly}} by our converged Ghost-GA solution
$\ket{\Psi_G}=\proj\ket{\Psi_0}$. 

Let us assume that among the $\tilde{\nu}$ Fermionic degrees of freedom 
in $\h$, see Eq.~\eqref{hubb-supp}, only $\nu$ are physical, while the other
$\tilde{\nu}-\nu$ degrees of freedom are ghost modes.
In other words, $\h$ is constructed utilizing only the
\emph{physical} ladder operators $\{\cc_{R\alpha},\ca_{R\alpha}\,|\,\alpha\in\{1,...,\nu\}\}$
(which implies, in particular, that 
$[\epsilon_{k}]_{\alpha\beta}=0\,\forall\,k,\forall\,\alpha,\beta>\nu$).

It is convenient to represent the Gutzwiller local maps $\projR$ as follows:
\be
\projR = \sum_{A,n=0}^{2^{\tilde{\nu}}-1} \tilde{\Lambda}_{A n}\,
\ket{A,R}\bra{n,R}
=\sum_{P=0}^{2^{{\nu}}-1}\,
\sum_{P'=0}^{2^{\tilde{\nu}-\nu}-1}\,
\sum_{n=0}^{2^{\tilde{\nu}}-1}
\tilde{\Lambda}_{(P,P') n}\,
\ket{(P,P'),R}\bra{n,R}
\label{Lam2-supp}
\,,
\ee
where 
\be
\ket{(P,P'),R}=[\cc_{R1}]^{q_1(P)}
\,...\;
[\cc_{R\nu}]^{q_{\nu}(P)}
\,
[\cc_{R\nu+1}]^{q_{1}(P')}
\,...\;
[\cc_{R,\tilde{\nu}}]^{q_{\tilde{\nu}}(P')}\,\ket{0}\,.
\ee
As we are going to show, since $\h$ depends only on the physical
degrees of freedom, i.e., it is constructed utilizing only the
\emph{physical} operators $\{\cc_{R\alpha},\ca_{R\alpha}\,|\,\alpha\in\{1,...,\nu\}\}$,
the coefficients defining the Gutzwiller projector are
of the form:
\be
\tilde{\Lambda}_{(P,P') n}=\Lambda_{Pn}\,\xi_{P'}\,.
\label{disentangled-supp}
\ee
Consequently, the converged Ghost-GA solution
$\ket{\Psi_G}=\proj\ket{\Psi_0}$ can be represented as
\be
\ket{\Psi_G}=\ket{\Psi^{phys}_G}\otimes\ket{\Psi^{ghost}_G}\,,
\label{DISENTANGLED-supp}
\ee
where $\ket{\Psi^{phys}_G}$, which resides entirely within the physical
subsystem, and $\ket{\Psi^{ghost}_G}$, which resides entirely within the
ghost subsystem, are \emph{disentangled}.

Let us now demonstrate Eq.~\eqref{disentangled-supp}.
Since we assumed that $[\epsilon_{k}]_{\alpha\beta}=0\,\forall\,k,\forall\,\alpha,\beta>\nu$,
from Eq.~\eqref{l3-supp} it follows that
\be
\D_{a\alpha}=0\quad\forall\,\alpha>\nu\,.
\label{d0-supp}
\ee
Consequently, the ground state $\ket{\Phi}$
of $\h_{\text{emb}}$, see Eq.~\eqref{hemb-supp},
is such that the subsystem generated by the ghost
degrees of freedom $\{\cc_{R\alpha},\ca_{R\alpha}\,|\,\alpha>\nu\}$
is \emph{disentangled} from the rest of the embedding system.
Thus, from Eq.~\eqref{pure-emb-supp} it follows that the matrix $\tilde{\phi}$
can be represented as
\be
\tilde{\phi}_{(P,P') n}=\phi_{Pn}\,\xi_{P'}\,.
\label{ppp-supp}
\ee 
The proof of Eq.~\eqref{disentangled-supp} (and, in turn, 
of Eq.~\eqref{DISENTANGLED-supp})
follows immediately from Eqs.~\eqref{ppp-supp} and \eqref{pl-supp}.

In summary, we have shown that
the Ghost-GA solution, see Eq.~\eqref{GWF-supp}, does not have any
spurious entanglement with the auxiliary ghost degrees of freedom
(as expected).

\subsection{F.~~~Exploiting Eq.~\eqref{DISENTANGLED-supp} numerically}

Since we know a-priori that Eq.~\eqref{DISENTANGLED-supp} must be satisfied by the
converged result, it is computationally convenient to impose from the onset
this condition onto the Ghost-GA variational parameters.

\begin{itemize}
	
	\item The first simplification due to Eq.~\eqref{DISENTANGLED-supp}
	arises from the fact that from Eq.~\eqref{d0-supp} implies that
	\be
	\Av{\Phi}{\hcc_{\alpha}\hfa_{a}}=0\quad\forall\,\alpha>\nu\,.
	\ee
	Consequently, from Eq.~\eqref{l6-supp} it follows that
	\be
	\R_{a\alpha}=0\quad\forall\,\alpha>\nu\,.
	\label{slice-supp}
	\ee
	This equation enables us to reduce substantially the computational complexity of 
	the root problem [Eq.~\eqref{roothubb-supp}], as the dimension of the space
	of matrices satisfying Eq.~\eqref{slice-supp} has only dimension
	$\nu\tilde{\nu}$ instead of $\tilde{\nu}^2$, see Eq.~\eqref{R-r-supp}.
	
	\item The second simplification arises from the fact
	that, because of Eq.~\eqref{d0-supp},
	the embedding Hamiltonian  
	[Eq.~\eqref{hemb-supp}]
	is such that its subsystem generated by 
	$\{\hcc_{R\alpha},\hca_{R\alpha}\,|\,\alpha\in\{1,...,\nu\}\}  \cup\{\hfc_{Ra},\hfa_{Ra}\,|\,a\in\{1,...,\tilde{\nu}\}\}$
	is \emph{disentangled} from 
	the ghost
	degrees of freedom $\{\hcc_{R\alpha},\hca_{R\alpha}\,|\,\alpha>\nu\}$.
	Consequently,
	it can be solved independently.
	
\end{itemize}

These observations reduce exponentially the scaling of the
computational complexity of the problem as a function of the number
of ghost orbitals used in the calculation.

\subsection{G.~~~Similarities with theory of Matrix Product States (MPS) and
	Projected Entangled Pair States (PEPS)}

In the previous subsections we have shown that the Ghost-GA wavefunction
$\ket{\Psi_G}=\proj\ket{\Psi_0}$ is disentangled from the auxiliary
subsystem. 
However, as we pointed out in the main text,
$\ket{\Psi_0}$ resides within the entire
Hilbert space --- including the ghost degrees of freedom.
In other words, the Gutzwiller operator $\proj=\prod_{R}\projR$
maps $\ket{\Psi_0}$ --- which is entangled with the auxiliary subsystem ---
into the a disentangled state represented as in Eq.~\eqref{DISENTANGLED-supp}.
It is interesting to note that,
in this respect, 
the Ghost-GA variational framework is very similar to
the theory of Matrix Product States (MPS) and
Projected Entangled Pair States (PEPS)~[\onlinecite{Vaestrate_DMRG-MPS-supp,Ulrich_review-supp}]. 
%

As we are going to show,
the main differences between the Ghost-GA theory and MPS/PEPS 
are the following:
\begin{itemize}
	\item Within the Ghost-GA theory,
	the local maps (i.e., the Gutzwiller projectors $\projR$)
	act on a generic "virtual" Slater determinant $\ket{\Psi_0}$ 
	belonging to an
	extended Hilbert space, which is determined variationally.
	\item Instead, MPS/PEPS
	are variational techniques formulated by applying local maps
	to fixed "pair states", which are also
	constructed within an extended Hilbert space~[\onlinecite{Vaestrate_DMRG-MPS-supp}].
\end{itemize}
In this respect, the variational ansatz represented in Eq.~\eqref{GWF-supp}
constitutes an extension of the MPS/PEPS variational space.
Of course, as we pointed out above and in the main text,
in our work we have also assumed the
Gutzwiller Approximation~[\onlinecite{Gutzwiller3-supp}]
(which is exact only in the limit of infinite dimensions) and 
the Gutziller constraints, see Eq.~\eqref{gconstr-supp} --- which are approximations
not employed in MPS/PEPS theory.
Because of this reason, while our approach constitutes a variationally
improved scheme with respect to the ordinary GA, it
is not "numerically exact".

In order to illustrate more clearly the connection between the Ghost-GA theory
and MPS/PEPS, let us consider a generic 1-dimensional Fermionic system
belonging to a Hilbert space represented as 
$\hilb := \bigotimes_{R=1}^\Omega \hilb_R$, where each local subsystem
$\hilb_R$ is generated by a set of Fermionic operators $\{\fc_{R\alpha}\}$,
where $a\in\{1,...,{\tilde{\nu}}\}$.
In this subsection we are going to assume also that $\tilde{\nu}$ is even.
Following, e.g., Ref.~[\onlinecite{Vaestrate_DMRG-MPS-supp}],
let us consider the following state:
\be
\ket{\Psi_0}=\bigotimes_{R=1}^\Omega \ket{B_{R}}\,,
\label{psi0-1d-supp}
\ee
which is a tensor product of maximally entangled "bonds" represented as
\be
\ket{B_{R}}=
\prod_{a=1}^{\frac{\bar{\nu}}{2}}
\frac{1}{\sqrt{2}}
\left[
\fc_{Ra}
+
\fc_{R+1\, a+\frac{\tilde{\nu}}{2}}
\right]
\,\ket{0}
\,,
\label{cool3-supp}
\ee
where $\fc_{R\,\Omega+1}=\fc_{R1}$.
Note that the state $\ket{\Psi_0}$ defined in
Eq.~\eqref{psi0-1d-supp} is a Slater determinant.
\begin{figure}
	\begin{center}
		\includegraphics[width=10.4cm]{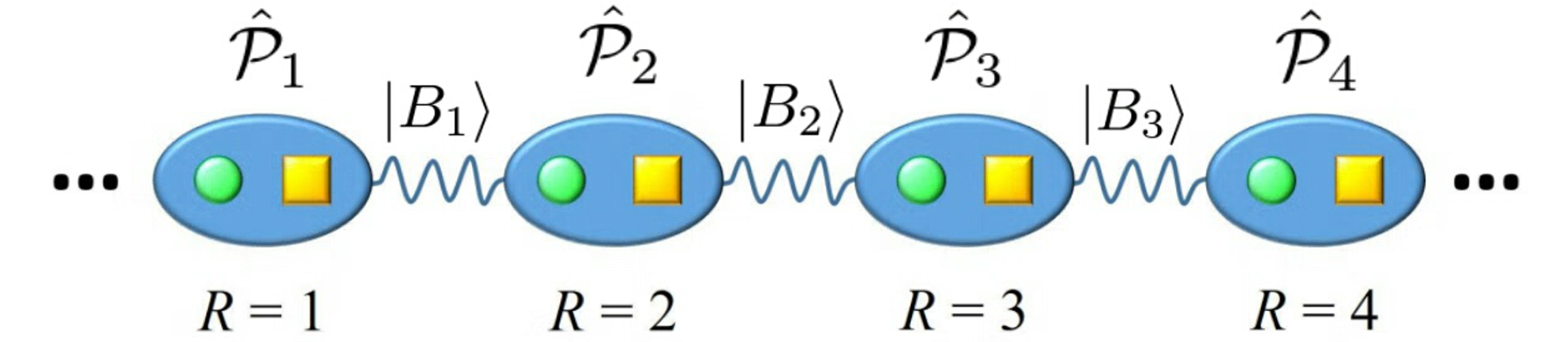}
		\caption{Construction of a MPS as a specific Ghost-Gutzwiller state.
		}
		\label{bonds-supp}
	\end{center}
\end{figure}

Let us now assume that only $\nu<\frac{\tilde{\nu}}{2}$
of the degrees of freedom are physical,
i.e., that we are interested in the ground state of a
Hamiltonian $\h$ constructed only with the
operators $\{\cc_{R\alpha},\ca_{R\alpha}\,|\,\alpha\in\{1,...,\nu\}\}$,
which are connected to the operators
$\{\fc_{Ra},\fa_{Ra}\,|\,a\in\{1,...,\tilde{\nu}\}\}$ by unitary transformations
\be
\cc_{R\alpha}=\sum_{a=1}^{\tilde{\nu}} [u_{R}]_{a\alpha}\,\fc_{Ra}
\,.
\ee
As in Sec.~I~E,
in order to construct a variational approximation of the ground state of
$\h$, we apply on $\ket{\Psi_0}$ a Gutzwiller operator
$\proj=\prod_R\projR$,
see Fig.~\ref{bonds-supp},
where $\projR$ is the most general map acting
over all of the local degrees of freedom labeled by $R$ represented as follows:
\be
\projR = \sum_{A,n=0}^{2^{\tilde{\nu}}-1} \tilde{\Lambda}_{A n}\,
\ket{A,R}\bra{n,R}
=\sum_{P=0}^{2^{{\nu}}-1}\,
\sum_{P'=0}^{2^{\tilde{\nu}-\nu}-1}\,
\sum_{n=0}^{2^{\tilde{\nu}}-1}
\tilde{\Lambda}_{(P,P') n}\,
\ket{(P,P'),R}\bra{n,R}
\label{Lam2-1d-supp}
\,,
\ee
where 
\be
\ket{(P,P'),R}=[\cc_{R1}]^{q_1(P)}
\,...\;
[\cc_{R\nu}]^{q_{\nu}(P)}
\,
[\cc_{R\nu+1}]^{q_{1}(P')}
\,...\;
[\cc_{R,\tilde{\nu}}]^{q_{\tilde{\nu}}(P')}\,\ket{0}\,,
\ee
and
the coefficients defining the Gutzwiller projector are
of the form:
\be
\tilde{\Lambda}_{(P,P') n}=\Lambda_{Pn}\,\xi_{P'}\,.
\label{disentangled-1d-supp}
\ee
Following Ref.~[\onlinecite{Vaestrate_DMRG-MPS-supp}],
it can be readily verified by inspection that the so obtained state
$\ket{\Psi_G}=\proj\ket{\Psi_0}$ can be represented as follows:
\be
\ket{\Psi_G}=\ket{\Psi^{phys}_G}\otimes\ket{\Psi^{ghost}_G}\,,
\label{DISENTANGLED-1d-supp}
\ee
where $\ket{\Psi^{phys}_G}$ is a MPS with bond dimension $2^{\tilde{\nu}}$,
which resides entirely within the physical subsystem.

In particular, the observation above clarifies that, within the Ghost-GA
variational ansatz,
increasing the number of ghost orbitals amounts essentially to increase
the so-called "bond dimension"~[\onlinecite{Vaestrate_DMRG-MPS-supp}],
which is the parameter controlling the accuracy of the variational 
ansatz also in MPS/PEPS.
However, we remark that
within the Ghost-GA theory the state $\ket{\Psi_0}$
is variationally determined, i.e., it is not restricted to the form [Eq.~\eqref{cool3-supp}].
In future works, it would be interesting to apply the
ghost-orbital Hilbert space extension without employing 
the Gutzwiller constraints and the Gutzwiller approximations, e.g., utilizing 
the Variational Monte Carlo method~[\onlinecite{VMC-Fermions-supp}].

\section{II.~~~Ghost-GA excitations from variational parameters}

\subsection{A.~~~Gutzwiller excitations and ARPES spectra in the standard multi-orbital GA}

Let us assume that
the Gutzwiller energy minimum is realized by a solution of 
Eqs.~\eqref{l2-supp}-\eqref{l7-supp} identified by the variational parameters
$\ket{\Psi_0}, E, \ket{\Phi},E^c, \R,\lambda, \D, \lambda^{c},\Delta$.
Thus, $\ket{\Psi_0}$ is the
ground state of $\Hqp$ corresponding to
the parameters $\R,\lambda$, see Eq.~\eqref{hqp-supp}.
Such a solution corresponds to a Gutzwiller wavefunction
which can be represented as
\be
\ket{\Psi_G}=\proj\ket{\Psi_0}\,,
\label{GWF-2-supp}
\ee
see Eq.~\eqref{GWF-supp}.
%
It is important to
observe that, in the thermodynamical limit
$\mathcal{N}\rightarrow\infty$,
Eqs.~\eqref{l2-supp}-\eqref{l7-supp} are satisfied also by
$\psi^\dagger_{kn}\ket{\Psi_0}, E+{\epsilon}^*_{kn}, \ket{\Phi},E^c, \R,\lambda, \D, \lambda^{c},\Delta$,
where $\psi^\dagger_{kn}$ is any eigenoperator of $\Hqp[\R,\lambda]$,
see Eq.~\eqref{hqp-supp}, and ${\epsilon}^*_{kn}$ is the corresponding
eigenvalue. Each one of the corresponding Gutzwiller states can
be represented as
\be
\ket{\Psi^p_{Gkn}}=\proj\,\psi^\dagger_{kn}\ket{\Psi_0}\,.
\label{GWF-QP-supp}
\ee
Similarly, Eqs.~\eqref{l2-supp}-\eqref{l7-supp} are satisfied also by
$\psi^\dagga_{kn}\ket{\Psi_0}, E-{\epsilon}^*_{kn}, \ket{\Phi},E^c, \R,\lambda, \D, \lambda^{c},\Delta$, which correspond to Gutzwiller
states represented as
\be
\ket{\Psi^h_{Gkn}}=\proj\,\psi^\dagga_{kn}\ket{\Psi_0}\,.
\label{GWF-QP-2-supp}
\ee

Since the states [Eqs.~\eqref{GWF-QP-supp} and \eqref{GWF-QP-2-supp}]
correspond to saddle
points of the Gutzwiller energy functional, they can be interpreted
as variational approximations
to many-body excitations of the system (with a number
of particles differing by 1 with respect to the ground state).
It is important to note that each one of these states
is connected to a single-particle excitation of $\Hqp$
through the action of the many-body operator $\proj$.
Thus, these states represent complex \emph{collective}
excitations of the system, which are commonly
called "Gutzwiller-Landau quasiparticles"~[\onlinecite{Gebhard-FL-supp}].

As pointed out in previous publications, see, e.g., Ref.~[\onlinecite{mythesis-supp}]
and the references therein,
this information allows us to evaluate the Gutzwiller ARPES spectra,
which is defined as follows:
\be
A_{\alpha\beta}(k,\omega)= \Av{\Psi_G}{\ca_{k\alpha}\,\delta(\omega-\h)\,\cc_{k\beta}}+
\Av{\Psi_G}{\cc_{k\beta}\,\delta(\omega+\h)\,\ca_{k\alpha}}\,.
\label{Ak-supp}
\ee
The main idea in order to evaluate approximately Eq.~\eqref{Ak-supp}
consists in inserting 
projectors over the Gutzwiller-Landau
quasiparticle states
$\sum_{n}\ket{\Psi^p_{Gkn}}\bra{\Psi^p_{Gkn}}$ and
$\sum_{n}\ket{\Psi^h_{Gkn}}\bra{\Psi^h_{Gkn}}$
(see Eqs.~\eqref{GWF-QP-supp} and \eqref{GWF-QP-2-supp})
in Eq.~\eqref{Ak-supp} and use the following identities:
\bea
\Av{\Psi_0}{\psi_{kn}\,\proj^\dagger\,\cc_{k\beta}\,\proj}&=&
\sum_{b}\R_{b\beta}\,\Av{\Psi_0}{\psi_{kn}\,\fc_{k\beta}}
\\
\Av{\Psi_0}{\proj^\dagger\,\ca_{k\alpha}\,\proj\,\psi^\dagger_{kn}}&=&
\sum_{a}\R^\dagger_{\alpha a}\,\Av{\Psi_0}{\fa_{ka}\,\psi^\dagger_{kn}}\,,
\eea
where $\fa_{k a}$ are the operators appearing in Eq.~\eqref{hqp-supp}.
From the identities above one can readily obtain~[\onlinecite{mythesis-supp}]:
\bea
A_{\alpha\beta}(k,\omega)&\simeq&
\sum_{nm}
\Av{\Psi_G}{\ca_{k\alpha}\,\ket{\Psi^p_{Gkn}}\bra{\Psi^p_{Gkn}}\,\delta(\omega-\h)\,\ket{\Psi^p_{Gkm}}\bra{\Psi^p_{Gkm}}\,\cc_{k\beta}}
\nonumber\\&+&
\sum_{nm}
\Av{\Psi_G}{\cc_{k\alpha}\,\ket{\Psi^h_{Gkn}}\bra{\Psi^h_{Gkn}}\,\delta(\omega+\h)\,\ket{\Psi^h_{Gkm}}\bra{\Psi^h_{Gkm}}\,\ca_{k\beta}}
\nonumber\\
&=&\big[\R^\dagger\,\delta(\omega-[\R {\epsilon}_{k}
\R^\dagger+\lambda])\,\R\big]_{\alpha\beta}\,.
\label{A-GA-supp}
\eea

Note that, in general, [Eqs~\eqref{GWF-QP-2-supp} and \eqref{GWF-QP-2-supp}] 
are not a complete basis. In fact, we have that, within the approximations
employed in Eq.~\eqref{A-GA-supp},
\be
\int_{-\infty}^{\infty}d\omega\, A_{\alpha\beta}(k,\omega)=[\R^\dagger\R]_{\alpha\beta}=Z_{\alpha\beta}
\label{defZ-supp}
\,.
\ee

Within the standard GA theory, the
interpretation of $Z$ as the matrix of quasiparticle weights is motivated by
the fact that the Green's function corresponding to Eq.~\eqref{A-GA-supp} is
\be
G_{\alpha\beta}(\omega)=\int_{-\infty}^{\infty}d\epsilon\,
\frac{A_{\alpha\beta}(k,\omega)}{\omega-\epsilon}
=\left[
\R^\dagger\,\frac{1}{\omega-[\R {\epsilon}_{k}
	\R^\dagger+\lambda]}\,\R
\right]_{\alpha\beta}=
\left[
\frac{1}{\omega-\epsilon_k-\Sigma(\omega)}
\right]_{\alpha\beta}\,.
\label{ggg-supp}
\ee
Thus, \emph{if $\R$ is invertible},
it can be straightforwardly verified that:
\be
{\Sigma}(\omega) = - \omega\,\frac{1-Z}{Z}
+\frac{1}{\R}\, \lambda\, \frac{1}{\R^{\dagger}}\,.
\label{linear-supp}
\ee
%
It is very important to note that 
Eq.~\eqref{linear-supp}, which provides a linear expression for the
self-energy as a function of the
frequency $\omega$, is valid \emph{only} if $\R$ is invertible.
However, we already know that this condition
is \emph{not} satisfied within the Ghost-GA theory, see Eq.~\eqref{slice-supp}.
This important point will be further stressed in the next subsection.


\subsection{B.~~~Gutzwiller excitations and ARPES spectra in the Ghost-GA theory}

As explained in the main text and in the previous section of the present supplemental material,
the Ghost-GA theory consists in applying the ordinary
multi-orbital Gutzwiller theory to the Hamiltonian "embedded" within
the extended Hilbert space obtained by introducing auxiliary ghost orbitals.
The only difference with respect to the ordinary theory is that,
as shown in the main text, utilizing the enlarged Gutzwiller
variational space leads to a better approximation to the ground
state $\ket{\Psi_G}$ and to the excited states
$\ket{\Psi^p_{Gkn}}$ and $\ket{\Psi^h_{Gkn}}$,
see Eqs.~\eqref{GWF-QP-supp} and \eqref{GWF-QP-2-supp}, which are all saddle points
of the Ghost-GA energy functional.

Let us consider again the Hamiltonian $\h$, see Eq.~\eqref{hubb-supp},
embedded within an enlarged Hilbert space.
In other words, we assume that $\h$ is constructed utilizing only the
\emph{physical} ladder operators
$\{\cc_{R\alpha},\ca_{R\alpha}\,|\,\alpha\in\{1,...,\nu\}\}$
(which implies, in particular, that 
$[\epsilon_{k}]_{\alpha\beta}=0\,\forall\,k,\forall\,\alpha,\beta>\nu$).

The Ghost-GA spectral function of the system is given
by the components of the spectral function
\be
A_{\alpha\beta}(k,\omega)=
\Av{\Psi_G}{\ca_{k \alpha}\,\delta(\omega-\h)\,\cc_{k \beta}}+
\Av{\Psi_G}{\cc_{k \beta}\,\delta(\omega+\h)\,\ca_{k \alpha}}
\label{Ghost-Ak-supp}
\ee
such that $\alpha,\beta\in\{1,...,\nu\}$.
The same steps utilized in the previous section lead to the
following approximation to the physical Green's function:
\be
G_{\alpha\beta}(\omega)=\int_{-\infty}^{\infty}d\epsilon\,
\frac{A_{\alpha\beta}(k,\omega)}{\omega-\epsilon}
\simeq\left[
\tR^\dagger\,\frac{1}{\omega-[\tR \tilde{\epsilon}_{k}
	\tR^\dagger+\tlambda]}\,\tR
\right]_{\alpha\beta}
\,,
\label{tG-supp}
\ee
Note also that, as expected, from Eqs.~\eqref{slice-supp} and \eqref{tG-supp}
it follows that
\be
G_{\alpha\beta}(\omega)=0\quad\forall\,\alpha,\beta>\nu\,,
\ee
i.e., only the physical components of the Green's function are non-zero.

The Ghost-GA physical self-energy is given, by definition, by
the Dyson equation, i.e., it is the function $\Sigma(\omega)$ satisfying
the following equation:
\be
G_{\alpha\beta}(\omega)
=\left[\frac{1}{\omega-{\epsilon}_k-{\Sigma}(\omega)}\right]_{\alpha\beta}
\quad\forall\,\alpha,\beta\leq\nu
\,,
\label{Ghost-Sigma-supp}
\ee
where $G_{\alpha\beta}(\omega)$ is given by Eq.~\eqref{tG-supp}.
It is important to note that, as pointed out above and in the main text,
$\R$ is \emph{not} invertible, see Eq.~\eqref{slice-supp}. 
Consequently, the Ghost-GA self-energy $\Sigma(\omega)$ defined above
is \emph{not} necessarily a linear function of the frequency $\omega$.

Remarkably, Eq.~\eqref{Ghost-Sigma-supp}
provides us with an \emph{analytical} expression for 
the Ghost-GA spectral function $A(k,\omega)$ in terms of
the bare physical dispersion $\epsilon_k$ and a few renormalization parameters,
i.e., the matrices $\tR$ and $\tlambda$.
Of course, the entries of $\tR$ and $\tlambda$ depend on the specific system
considered, as they
have to be calculated numerically from Eqs~\eqref{l2-supp}-\eqref{l7-supp}.
In particular, Eqs.~8-13 of the main text have been obtained from
Eq.~\eqref{Ghost-Sigma-supp}.

Note that, within the context of the Ghost-GA theory, the excited states
$\ket{\Psi^p_{Gkn}}$ and $\ket{\Psi^h_{Gkn}}$ are constructed
by applying the Gutzwiller projector to states
belonging to the extended Hilbert space. 
In this respect, these mathematical form of these
states present suggestive formal analogies with
the "hidden Fermi liquid" excitations
previously introduced by P. W. Anderson~[\onlinecite{Hidden-FL-supp,And-HFL-PNAS-supp}].

We point out also that the construction above is \emph{not} specific to the
single band Hubbard Hamiltonian considered in this work, as it can 
be straightforwardly applied to generic Hubbard Hamiltonian ---
with an arbitrary number of physical orbitals.

\section{III.~~~Additional Remarks about the benchmark calculations of the single band Hubbard model}

\subsection{A.~~~Behavior of the Ghost-GA self-energy}

\begin{figure}
	\begin{center}
		\includegraphics[width=10.4cm]{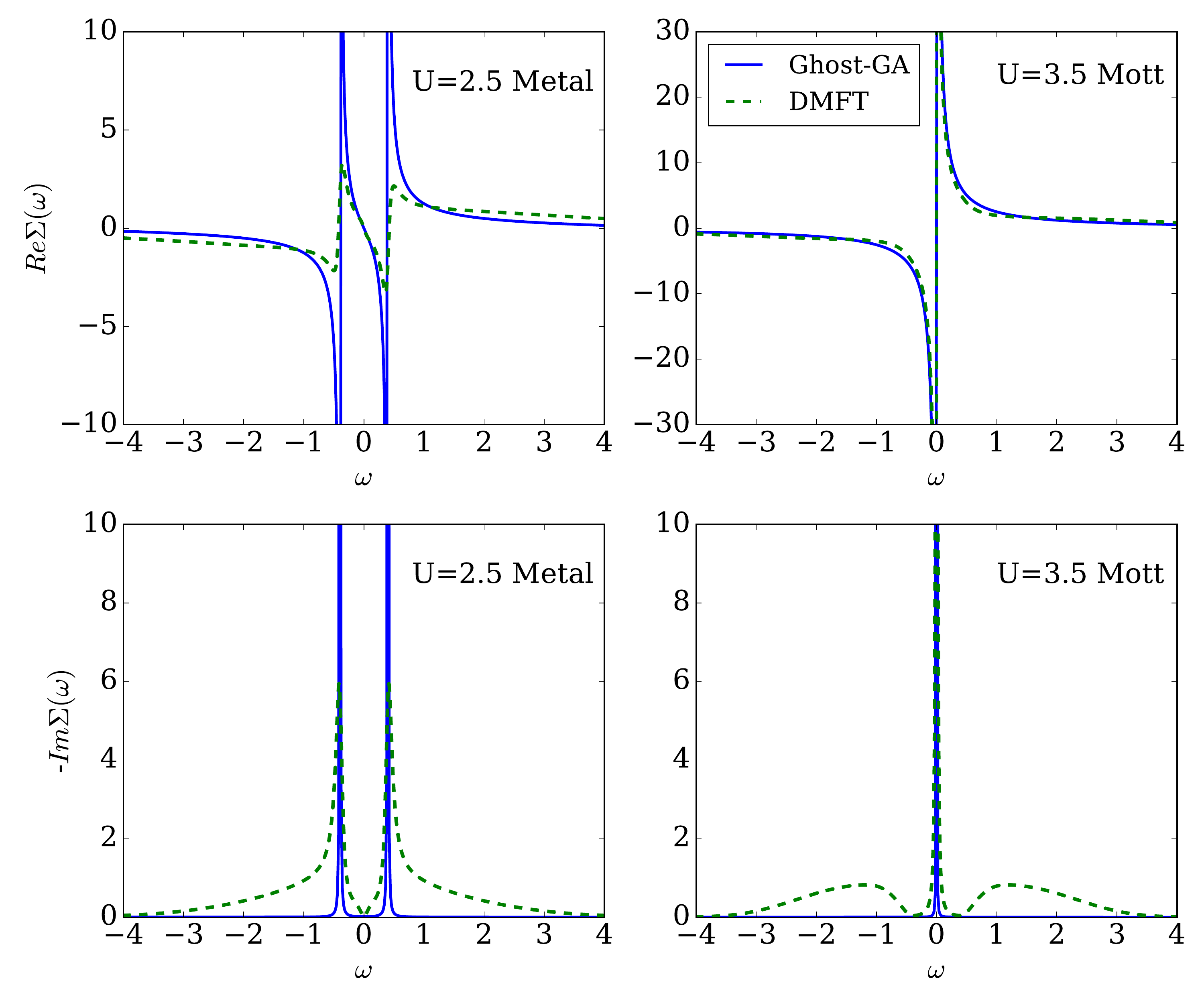}
		\caption{Behaviour of the Ghost-GA
			self-energy $\Sigma(\omega)$ obtained 
			utilizing 2 ghost orbitals at $U=2.5$ (metallic phase)
			and 1 ghost orbital at $U=3.5$ (Mott-insulator phase),
			in comparison with DMFT.
		}
		\label{sigma-comparison-supp}
	\end{center}
\end{figure}

As discussed in the main text, 
the Ghost-GA self-energy of the single band Hubbard Hamiltonian
can be represented as follows in the metallic phase:
\be
\Sigma(\omega)=\frac{\omega}{1+\frac{1}{z-
		\frac{\omega^2-l^2+2h\omega^2}{\omega^2-l^2}}
}
= -\frac{1-z}{z}\,\omega + o(\omega^2)
\,
\label{sigM-supp}
\ee
while it can be represented as follows in the Mott phase:
\be
\Sigma(\omega)=-\frac{1-h}{h}\,\omega + \frac{l^2}{h}\,\frac{1}{\omega}
\,.
\label{sigI-supp}
\ee
The values of the parameters $z$ and $h$ evolve
as illustrated in Fig.~2 of the main text.
The behavior of the Ghost-GA self-energy is shown in Fig.~\ref{sigma-comparison-supp}
in comparison with DMFT.

As discussed in the main text, some of the main features of the
self energy are captured by Eqs.~\eqref{sigM-supp} and \eqref{sigI-supp}.
The differences with respect to the DMFT solution arise mostly from the fact
that within the Ghost-GA theory, by construction, 
the self-energy can develop only poles, see Eqs.~\eqref{sigM-supp} and \eqref{sigI-supp},
but can not capture branch-cut singularities on the real axis.
In other words, the so-called "scattering rate" is
not captured by our approximation
(as it is not captured by the ordinary GA).

It is interesting to observe that in the limit of $U\rightarrow 0$
we have $z\rightarrow 1$ and $h\rightarrow 0$.
Thus, from Eq.~\eqref{sigM-supp} we deduce that
$\lim_{U\rightarrow 0}\Sigma(\omega)\rightarrow 0\; \forall\, \omega$,
i.e., the uncorrelated limit is described exactly by our theory.
Note that this result is to be expected as, by construction,
the Ghost-GA variational space includes all of the Slater determinants.

\subsection{B.~~~Ghost-GA solution utilizing 3 ghost orbitals}

In this work, the accuracy of the Ghost-GA theory
has been explicitly demonstrated 
(in all regimes of interaction strength)
on the single-band Hubbard Model
for a semi-circular density of states (DOS) ---
which corresponds, e.g.,
to the Bethe lattice in the limit of infinite coordination number,
where DMFT is \emph{exact}.
Specifically, the calculations discussed in the main text
were performed using 2 ghost orbitals
in the metallic phase and 1 ghost orbital in the Mott phase,
as this was the minimal ghost extension enabling us to capture
the main features of the ARPES spectra and of the $T=0$ phase diagram
of the system.
In this section we are going to discuss how the Ghost-GA result depends on the 
number of ghost orbitals used in the calculations.

\begin{figure}
	\begin{center}
		\includegraphics[width=10.4cm]{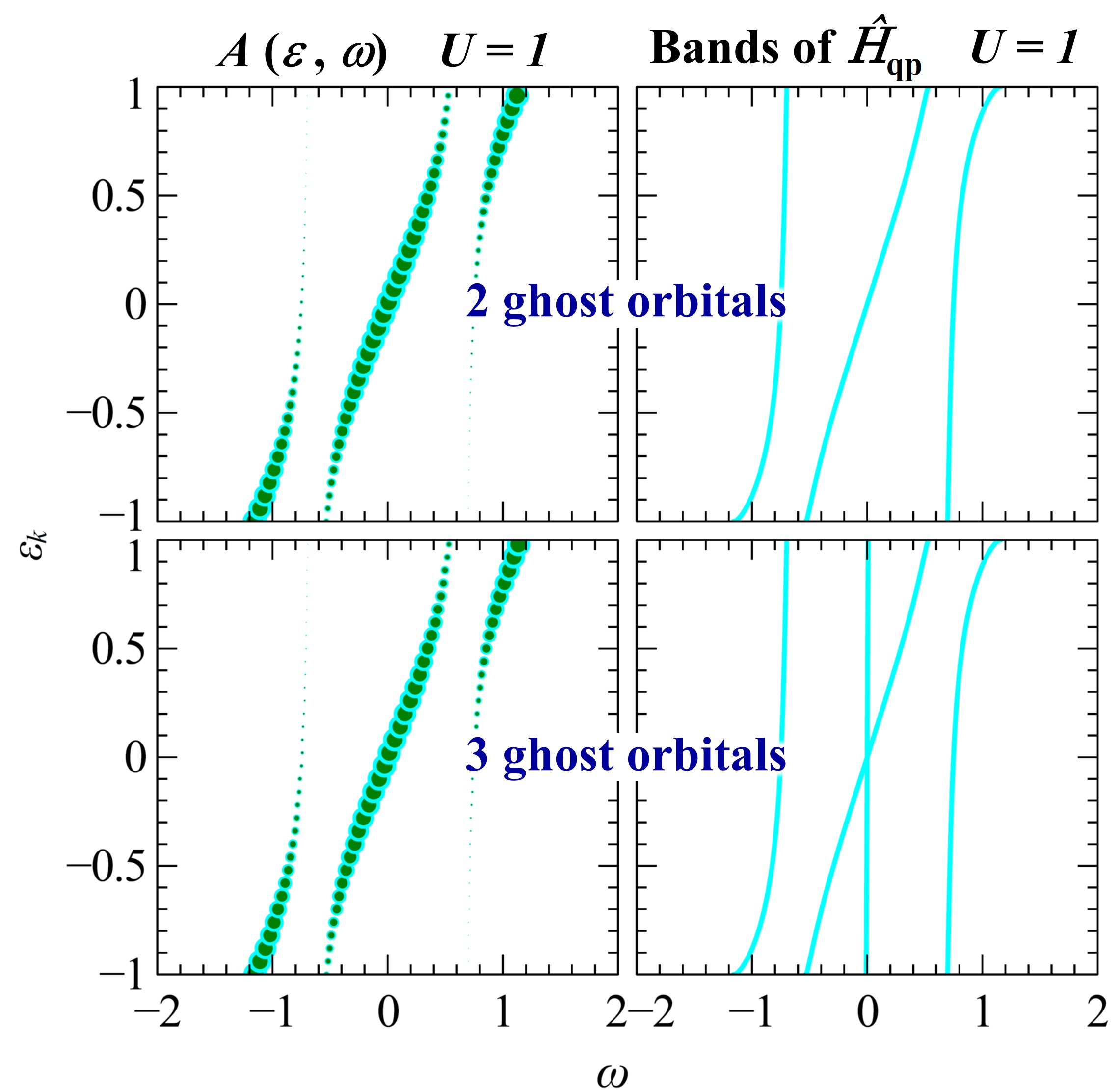}
		\caption{Poles of the physical Ghost-GA
			energy-resolved Green's function (left panels) 
			and bands of the Ghost-GA quasiparticle Hamiltonian obtained
			in the main text 
			utilizing 1 ghost orbital (upper panels) and
			3 ghost orbitals (lower panels)
			at $U=1$ (metallic phase).
			The size of the bullets indicates the spectral weights
			of the corresponding poles.
		}
		\label{figure1-supp}
		%
		\includegraphics[width=10.4cm]{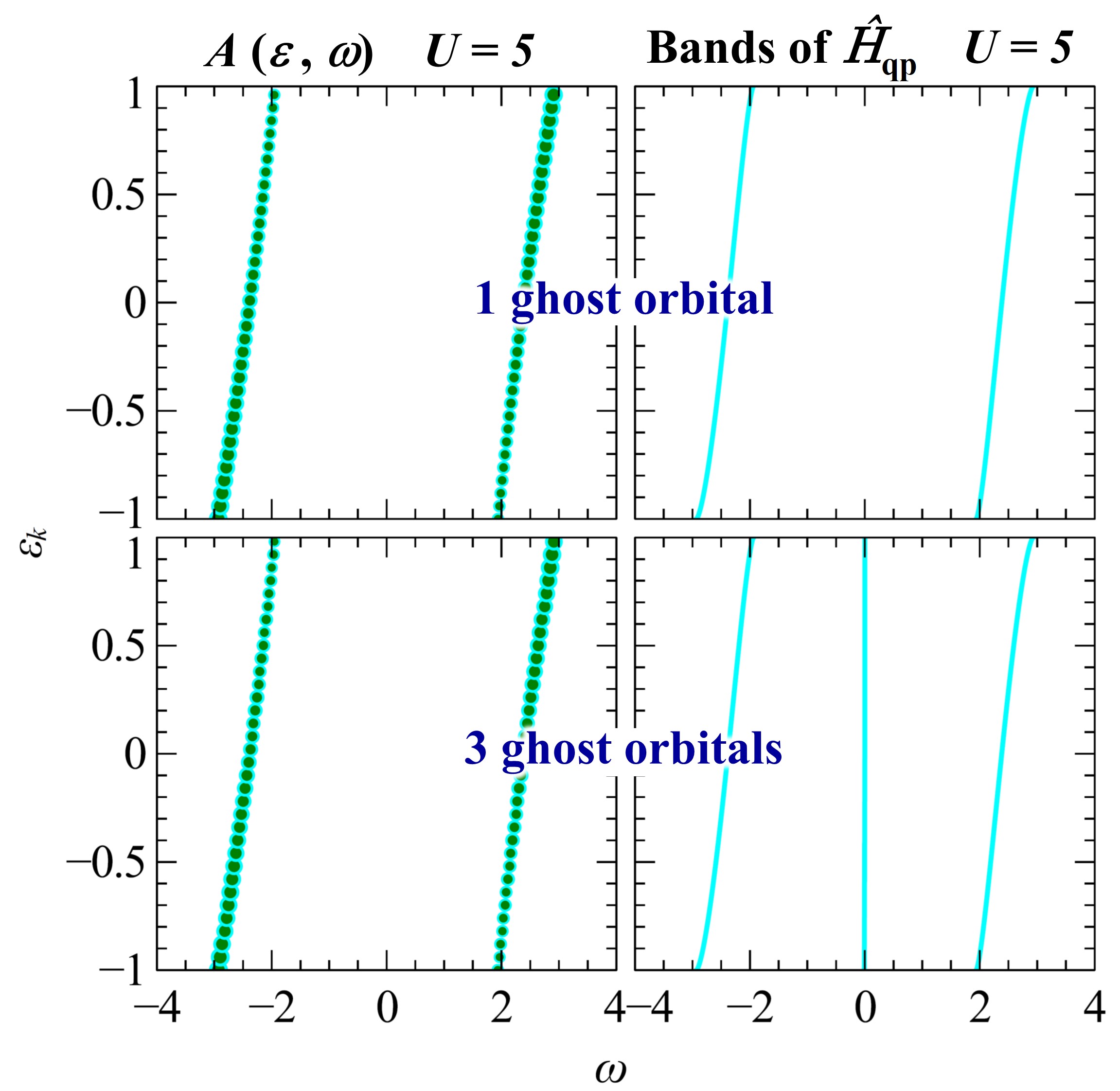}
		\caption{Poles of the physical Ghost-GA
			energy-resolved Green's function (left panels) 
			and bands of the Ghost-GA quasiparticle Hamiltonian obtained
			in the main text 
			utilizing 1 ghost orbital (upper panels) and
			3 ghost orbitals (lower panels)
			at $U=5$ (Mott phase).
			The size of the bullets indicates the spectral weights
			of the corresponding poles.
		}
		\label{figure2-supp}
	\end{center}
\end{figure}

The Ghost-GA physical Green's function of the single-band Hubbard model
\be
G(\omega)=\left[
\tR^\dagger\,\frac{1}{\omega-[\tR \tilde{\epsilon}_{k}
	\tR^\dagger+\tlambda]}\,\tR
\right]_{11}\,,
\label{tG-2-supp}
\ee
see the third member of Eq.~\eqref{tG-supp},
is analytical by construction everywhere over the upper-half plane,
and has poles over the eigenvalues $\tilde{\epsilon}^*_{kn}$ of: 
\be
\Hqp = 
\sum_{k\sigma}\sum_{ab} 
\big[\tR \tilde{\epsilon}_{k}
\tR^\dagger+\tlambda\big]_{ab}\,\fc_{k a\sigma}\fa_{k b\sigma}
= \sum_{k\sigma}\sum_{n} 
\tilde{\epsilon}^*_{kn}\,
\psi^\dagger_{kn\sigma}\psi^\dagga_{kn\sigma}\,.
\label{Ghost-hqp-supp}
\ee
However, the physical spectral weights of these poles is generally not $1$
because of the presence of the renormalization
coefficients $\tR$ in Eq.~\eqref{tG-2-supp}.

As mentioned in the main text, if 2 ghost orbitals are used to study 
the Mott phase of the Hubbard model, all of the results, including 
both the ground state properties and the ARPES spectra, remain essentially
unchanged.
In particular, even though $\Hqp$ has 3 bands (by construction), the converged
values of the renormalization parameters are such that one of these bands
is flat and acquires spectral weight $0$ (i.e., it does not contribute
to the ARPES spectra), while the other 2 bands
do not display any visible difference with respect to those computed
using only 1 ghost orbital.
Here we show that, as expected, the same mechanism occurs when 3 ghost orbitals
are used to describe the system, both in the metallic and in the
insulating phases of the single-band 
Hubbard model with semicircular DOS at half-filling.

In Figs.~\ref{figure1-supp} and \ref{figure2-supp}
the Ghost-GA energy-resolved spectral function
$A(\epsilon,\omega)=-\frac{1}{\pi}\text{Im}\,{G(\epsilon,\omega)}$,
see Eq.~\eqref{tG-2-supp} is analyzed for 2 values of $U$
in comparison with the Ghost-GA quasiparticle spectral function
$\Tr\big[\delta(\omega-[\tR \,\tilde{\epsilon}\, \tR^\dagger+\tlambda])\big]
=\sum_n \delta(\omega-\tilde{\epsilon}^*_n)$,
see Eq.~\eqref{Ghost-hqp-supp}.
%
The Ghost-GA results obtained in the main text
--- utilizing 
2 ghost orbitals in the metallic phase and 1 ghost orbital in
the Mott phase ---
are compared with the results obtained using 3 ghost orbitals for
both phases.

This analysis shows that, in all cases analyzed so far,
increasing the number of orbitals (ghost plus physical) 
beyond the minimal number necessary in order to match
the number of bands in the system
does not affect
the physical spectral function $\A(\epsilon,\omega)$.
In particular, in the results shown in Figs.~\ref{figure1-supp} and \ref{figure2-supp} 
we observe that the physical spectral function obtained with
3 ghost orbitals displays
1 flat band with zero spectral weight in the metallic phase,
while it displays 2 degenerate flat bands with 0 spectral weight
in the Mott phase.

\subsection{C.~~~Benchmark calculations of the single-band Hubbard model away from half-filling}

In order to study the Hubbard Hamiltonian away from half filling, here we work
in the grand-canonical ensemble, i.e., we study the following Hamiltonian:
\be
\sum_{k\sigma} 
{\epsilon}_{k}\,
\cc_{k1\sigma}\ca_{k1\sigma}
+\sum_{R}{U}\,
\hat{n}_{R 1\uparrow}\hat{n}_{R 1\downarrow} + 
\left(\mu-\frac{U}{2}\right)\,\sum_{R\sigma}
\hat{n}_{R 1\sigma}\,.
\label{muhubb-supp}
\ee
Note that Eq.~\eqref{muhubb-supp} is represented in such a way that
the particle-hole symmetric case
(half-filling) corresponds to $\mu=0$. 
\begin{figure}
	\begin{center}
		\includegraphics[width=18.4cm]{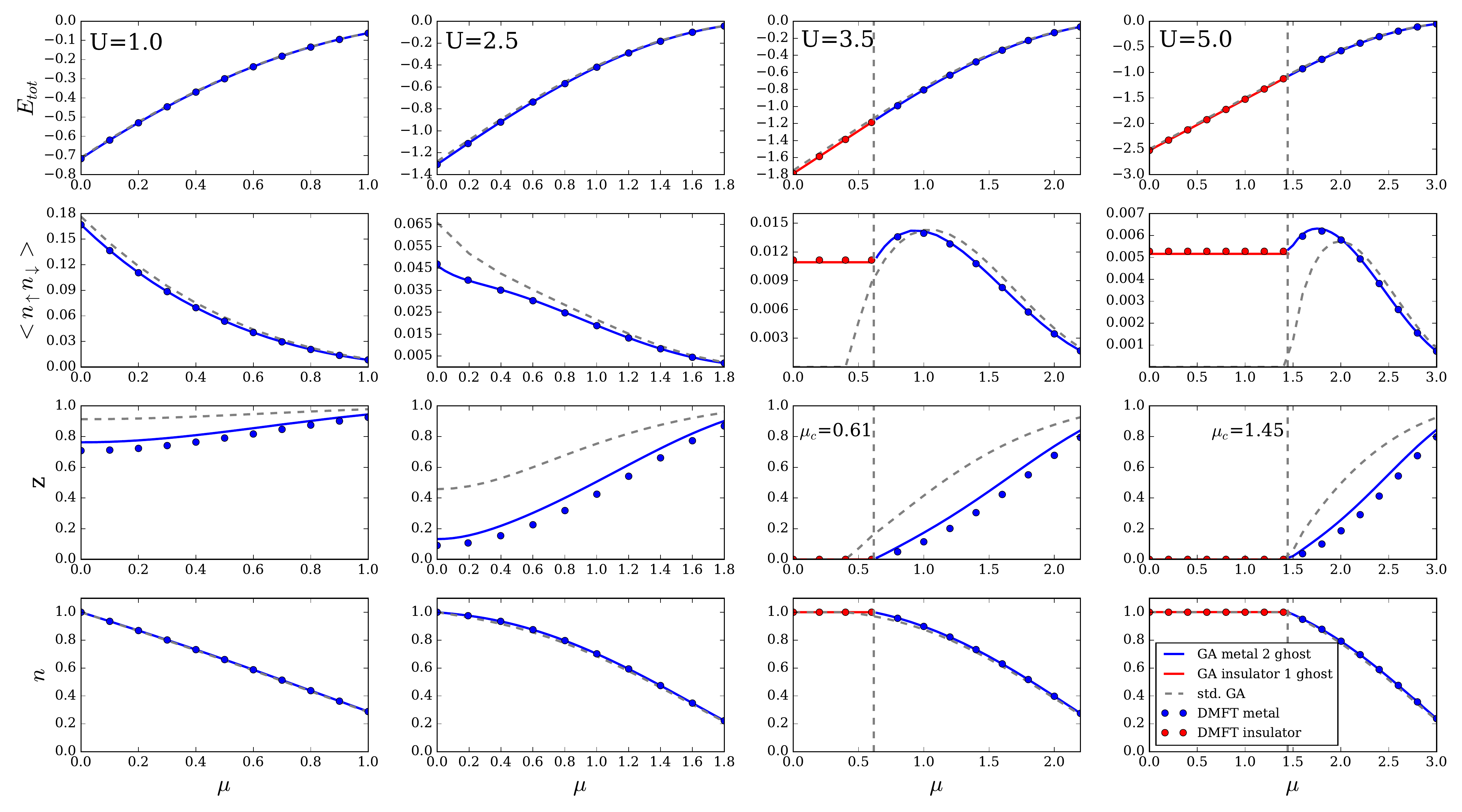}
		\caption{Evolution of 
			total energy, local double occupancy,
			QP weight and local occupancy 
			for the single-band Hubbard model
			with semicircular DOS
			as a function of
			the Hubbard chemical potential $\mu$, for different
			values of interaction strength $U$.
			The Ghost-GA results (blue curves) are shown in comparison with
			the ordinary GA and with the DMFT+NRG results.
		}
		\label{mu-supp}
	\end{center}
\end{figure}
In Fig.~\ref{mu-supp} is shown the evolution
as a function of the chemical potential $\mu$ of the
total energy, the local double occupancy, the QP weight $z$
and the local occupancy, for 2 values of the 
Hubbard interaction strength $U$.
The Ghost-GA results are shown
in comparison with the ordinary GA theory and with DMFT.

As in the case of half filling considered above and in the main text,
the agreement between Ghost-GA and DMFT
is quantitatively remarkable.

\section{IV.~~~Ghost-GA benchmark calculations of the Single-orbital Anderson Impurity Model}

In order to provide further evidence of the quality the Ghost-GA theory
and, in turn, to further validate the results provided in the main text,
in this section we provide also benchmark calculations of the single-band
Anderson Impurity Model (AIM).
\bea
\himp&=&\frac{U}{2}\Big[1-\sum_\sigma\cc_{0\sigma}\ca_{0\sigma}\Big]^2
+\sum_{\sigma=\pm \frac{1}{2}}\sum_{k} \left( V_{k}\,\cc_{k\sigma}\ca_{0\sigma}+\text{H.c.}\right)
+\sum_{\sigma=\pm \frac{1}{2}} \sum_{k} \epsilon_{k}\,\cc_{k\sigma}\ca_{k\sigma}
\,,
\label{singleimp-supp}
\eea
where $\sigma$ is the spin label.
In particular, we are going to consider the case of a 
flat density of states, whose hybridization function~[\onlinecite{Hewson-supp}]
can be represented as follows:
\be
\lim_{\eta\rightarrow 0^+} [\Delta(\omega+i\eta)]_{\sigma\sigma'}
= \delta_{\sigma\sigma'}\, \left[
\frac{\Gamma}{\pi}\ln\left|\frac{\omega + D}{\omega - D}\right|-i\Gamma\theta(D^{2}-\omega^{2})
\right]
\,,
\label{g_Gamma-supp}
\ee
where $\theta$ is the Heaviside step function.

%

%
\begin{figure}
	\begin{center}
		\includegraphics[width=8.4cm]{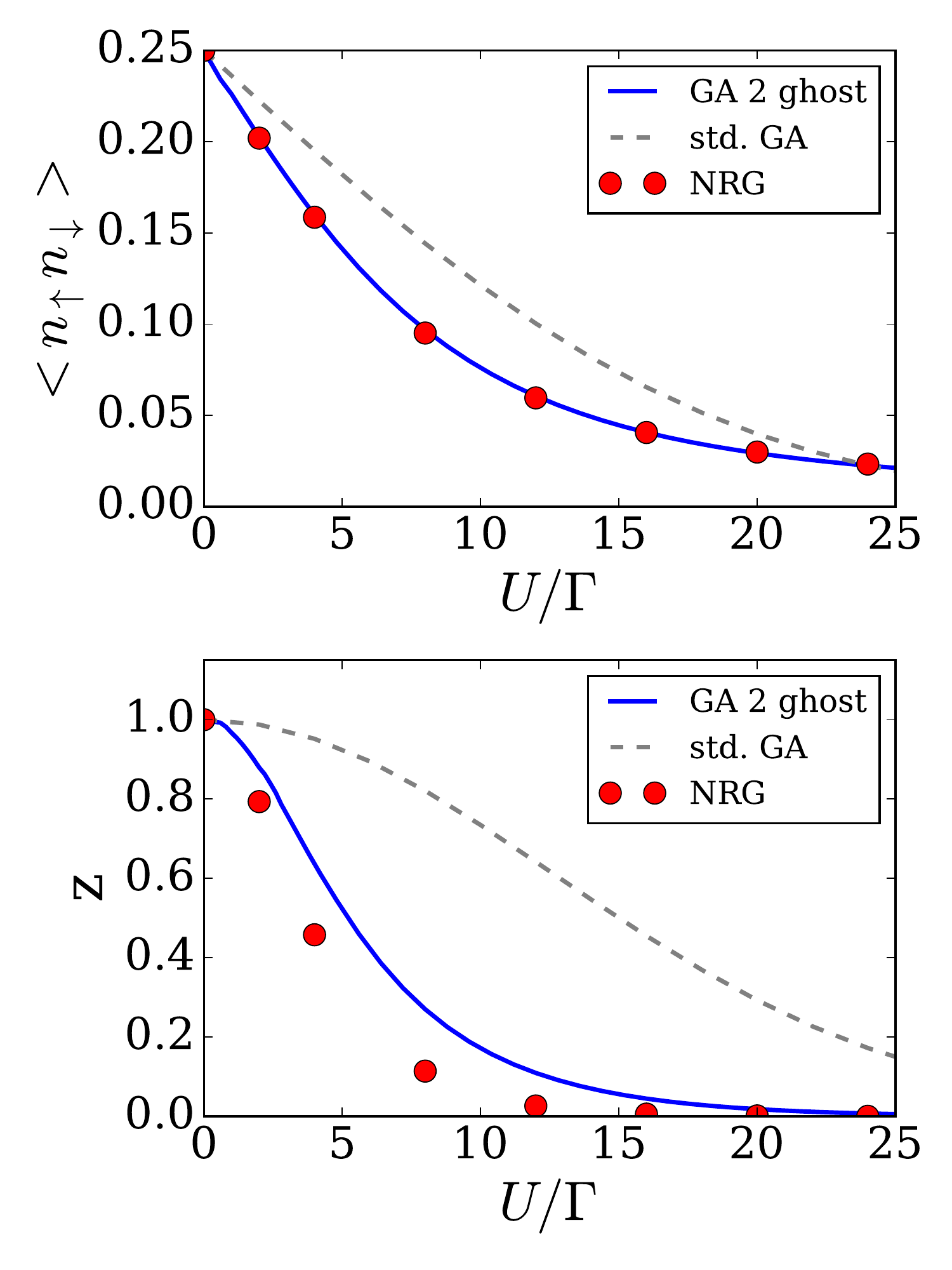}
		\caption{Evolution
			as a function of the Hubbard interaction strength $U$ of 
			the impurity double occupancy and the QP weight $z$.
			The Ghost-GA results obtained by adding 2 ghost orbitals on the impurity
			site (blue curves) are shown
			in comparison with the ordinary GA theory and with NRG.
		}
		\label{U_d_Z-supp}
	\end{center}
\end{figure}
In Fig.~\ref{U_d_Z-supp} is shown the evolution
as a function of the Hubbard interaction strength $U$ of 
the impurity double occupancy and the QP weight $z$.
The Ghost-GA results obtained by adding 2 ghost orbitals on the impurity
site are shown
in comparison with the ordinary GA theory and with NRG.

Consistently with the benchmark calculations of the
Hubbard model considered above
and in the main text, here we observe that the accuracy of the Ghost-GA 
solution is quantitatively remarkable also for the AIM.
In particular, we note that
the Ghost-GA results are substantially more accurate
with respect to the ordinary GA 
--- as expected, since
the Ghost-GA theory enlarges the ordinary Gutzwiller variational space.

It is interesting to analyze more in detail the behavior of $z$ (which is
commensurate with the Kondo temperature~[\onlinecite{Hewson-supp}]), in the strong-coupling regime
$U\gg \Gamma$.
It is well known that the ordinary Gutzwiller theory
provides, even \emph{without} ghost orbitals,
an exponentially decaying Kondo scale in the large-$U$ limit:
$z\sim e^{-\frac{\pi}{16} \frac{U}{\Gamma}}$.
The only well-known discrepancy of this result with respect to
the exact solution of the AIM
is that the Gutzwiller prefactor 
at the exponent is $\pi/16$, 
while the exact universal numerical value is $\pi/8$, see, e.g.,
Eqs.~101-102 of Ref.~[\onlinecite{metr-supp}] and the references therein.
A numerical analysis of the results displayed in Fig.~\ref{U_d_Z-supp}
shows that the Ghost-GA theory obtained by
adding 2 ghost orbitals provides, in particular,
a more accurate estimation of the
universal prefactor in front of $U/\Gamma$
at the exponent of the expression for $T_K$,
which behaves as
$z\sim e^{-\frac{\pi}{13.7} \frac{U}{\Gamma}}$ in the large-$U$ regime.

A more detailed study of the AIM will be provided in future publications.

\section{V.~~~Ghost Rotationally Invariant Slave Boson ($\text{Ghost-RISB}$) theory}

As mentioned in the conclusions of the main text, the Ghost-GA theory can
be equivalently reformulated as the mean field approximation of an
alternative formulation of the Rotationally Invariant Slave Boson (RISB) theory,
which we call Ghost-RISB theory.
For completeness, this construction is briefly described in this section.

Let us consider a generic Hubbard Hamiltonian
represented as in Eq.~\eqref{hubb-supp}.
%
In Ref.~[\onlinecite{Our-SB-supp}] it was demonstrated that 
Eq.~\eqref{hubb-supp} can be equivalently reformulated as a gauge theory
in an auxiliary Hilbert space (i.e., the RISB theory), and that
this exact reformulation of the many-body problem reduces to
the ordinary GA at the mean-field level.

Here we observe that the mapping mentioned above could be equivalently applied
to the Hubbard Hamiltonian expressed within a Ghost-GA
extended Hilbert space.
%
Of course, this construction would result in an alternative
\emph{exact} reformulation of
the many body problem, which we call Ghost-RISB theory.
However, by construction, the Ghost-RISB theory reduces to the Ghost-GA
theory at the mean field level --- which is substantially more accurate with
respect to the ordinary GA.
Because of this reason, it would be very interesting to explore the possibility
of taking into account fluctuations beyond mean-field within the
framework of the Ghost-RISB, as it has been done in previous works
within the ordinary SB theory~[\onlinecite{fluctuations-SB-2017-supp}].


\begin{thebibliography}{56}
\bibitem[{foo({\natexlab{a}})}]{footnote-3}
	\eprint{Note also that, as the MPS, the Ghost-GA wavefunction does not break
		artificially the translational invariance of the system.}
\bibitem[{foo({\natexlab{b}})}]{footnote-1}
	\eprint{Our results do not qualitatively depend on the specific choice of DOS.}
\bibitem[{foo({\natexlab{c}})}]{footnote-2}
	\eprint{Eq.~\eqref{g} is invariant with respect to any gauge transformation
		$\tR\rightarrow u^\dagger\,\tR$, $\tlambda\rightarrow
		u^\dagger\,\tlambda\,u$, where $u$ is unitary, see Ref.~\cite{Our-SB}.}
\bibitem[{foo({\natexlab{d}})}]{footnote-4}
	\eprint{Note that the expression for $\Sigma(\omega)$ is $k$ indepenent as
		expected, since we employed the GA}.
\end{thebibliography}
\end{document}